\documentclass[12pt,english,floatfix,nofootinbib,superscriptaddress,aps,prd,preprint]{revtex4}
\usepackage[utf8]{inputenc}
\usepackage{float}
\usepackage{array}
\usepackage{bbold}
\usepackage{lipsum}
\usepackage{dsfont}
\usepackage{graphicx}
\usepackage{amsmath,amsthm,amsfonts,amssymb}
\usepackage{graphicx}
\usepackage[english]{babel} 
\usepackage{color}
\usepackage{tensor}
\usepackage{esint}
\usepackage[dvips]{epsfig}
\usepackage[dvips]{graphicx}
\usepackage{float}
\usepackage{units}
\usepackage{textcomp}
\usepackage{mathrsfs}
\usepackage{amsmath}
\usepackage[makeroom]{cancel}
\usepackage{amssymb}
\usepackage{amsbsy}
\usepackage{amsfonts}
\usepackage{amssymb,mathrsfs,xcolor}
\usepackage{esint}
\usepackage{braket}
\usepackage{array}
\usepackage{graphicx}

\usepackage{wasysym}
\usepackage{multirow}
\usepackage{wrapfig}
\usepackage{subfig}

\usepackage{stmaryrd}
\usepackage{upgreek}

\makeatletter

\makeatletter\usepackage{babel}

\usepackage{hyperref}
\hypersetup{
    colorlinks,
    citecolor=blue,
    filecolor=green,
    linkcolor=purple,
    urlcolor=red,
}

\usepackage{slashed}

\newcommand{\ie}{\begin{equation}}
\newcommand{\fe}{\end{equation}}
\newcommand{\se}{\begin{eqnarray}}
\newcommand{\ff}{\end{eqnarray}}

\begin{document}

\title{Implications of a Simpson--Visser solution in Verlinde's framework}

\author{A. A. Ara\'{u}jo Filho}
\email{dilto@fisica.ufc.br}

\affiliation{Departamento de Física Teórica and IFIC, Centro Mixto Universidad de Valencia--CSIC. Universidad
de Valencia, Burjassot--46100, Valencia, Spain}

\affiliation{Departamento de Física, Universidade Federal da Paraíba, Caixa Postal 5008, 58051-970, João Pessoa, Paraíba, Brazil}


\date{\today}

\begin{abstract}

This study focuses on investigating a regular black hole within the framework of Verlinde's emergent gravity. In particular, we explore the main aspects of the modified Simpson--Visser solution. Our analysis reveals the presence of a unique physical event horizon under certain conditions. Moreover, we study the thermodynamic properties, including the \textit{Hawking} temperature, the entropy, and the heat capacity. Based on these quantities, our results indicate several phase transitions. Geodesic trajectories for photon--like particles, encompassing photon spheres and the formation of black hole shadows, are also calculated to comprehend the behavior of light in the vicinity of the black hole. Additionally, we also provide the calculation of the time delay and the deflection angle. Corroborating our results, we include an additional application in the context of high–-energy astrophysical phenomena: neutrino energy deposition. Finally, we investigate the quasinormal modes using third--order WKB approximation.

\end{abstract}

\maketitle


\section{Introduction}

General Relativity, while a remarkable theory, is acknowledged to be incomplete when it comes to describing the behavior of the universe at both the classical and quantum levels. One significant issue arises from the existence of singularities, which are problematic within this framework. Classical Einstein's theory of gravity encounters challenges when dealing with unavoidable singularities found in solutions like Schwarzschild, Reisner--Nordström, and Kerr metrics, which exhibit these peculiar features within their interiors. The scientific consensus recognizes the need for modifications to general relativity in regions where spacetime curvature becomes significantly high. These modifications are crucial for a more comprehensive understanding of gravity in extreme conditions.

In addition to addressing the presence of singularities, the modification of general relativity is essential for achieving a theory that is ultraviolet (UV) complete. Various proposals have emerged to accomplish this necessary modification. Extensive research has shown that incorporating higher--order curvature terms and terms involving higher derivatives can significantly improve the UV properties of Einstein's gravity \cite{4,2,1,3}.

However, a common challenge faced by these modified theories is the presence of non-physical degrees of freedom known as ghosts \cite{6,7,5}. Fortunately, in recent years, a groundbreaking UV--complete modification of general relativity has been proposed, effectively addressing this issue \cite{7,6,5}. Such an approach, known as ghost--free gravity \cite{5,6,7,9,8,14,13,12,10}, incorporates an infinite number of derivatives and exhibits fascinating non--local characteristics \cite{13,12,10}. Interestingly, a similar theory naturally emerges within the framework of non--commutative geometry deformation of Einstein's gravity \cite{16,15}, as thoroughly discussed in a comprehensive review and its referenced works \cite{17}. Moreover, the application of ghost--free gravity has been extensively explored in the context of studying singularities in cosmology and black holes \cite{19,18,20,21,23,22,25,24}.

In the absence of a specific theory, exploring potential modifications that could arise when gravity achieves UV completeness provides valuable insights. Such investigations are particularly informative when certain ``natural" assumptions about the properties of a comprehensive theory are taken into account. In this context, our focus centers on the exploration of regular (non--singular) models of black holes, aiming to investigate black hole metrics that lack curvature singularities. The pioneering work of Bardeen \cite{26} introduced the concept of a non--singular black hole, where the singularity was replaced by a charged matter core resulting from the collapse of charged matter. Additionally, a variety of models depicting such a feature have been proposed and discussed, including neutral, charged, and rotating configurations \cite{28,27,29,34,33,32,31,30,35,37,36,39,38,neves2014regular,frolov2016notes,maluf2018thermodynamics,neves2017deforming,maluf2019bardeen,neves2020accretion,40,neves2017bouncing}.

Nevertheless, a comprehensive grasp of gravitational waves and their properties is indispensable in exploring a myriad of physical phenomena, ranging from cosmological events in the primordial universe to astrophysical processes such as the evolution of stellar oscillations \cite{dziembowski1992effects,kjeldsen1994amplitudes,unno1979nonradial} and binary systems \cite{yakut2005evolution,hurley2002evolution,heuvel2011compact,pretorius2005evolution}. These waves exhibit a diverse spectrum of intensities and distinct characteristic modes, with their spectral traits profoundly influenced by the underlying phenomena that engender them \cite{riles2017recent}. When matter undergoes gravitational collapse, giving rise to the formation of a black hole, it enters a perturbed state, emitting radiation that encompasses an array of discrete frequencies unrelated to the collapse process itself \cite{konoplya2011quasinormal}. Termed \textit{quasinormal} modes, these perturbations exhibit distinct frequencies that define their unique nature \cite{kokkotas1999quasi,heidari2023gravitational}.

The investigation of \textit{quasinormal} modes of black holes has garnered extensive attention in the literature, employing the weak field approximation. This approach has been employed not only within the framework of general relativity \cite{franzin2022scalar,santos2016quasinormal,rincon2020greybody,berti2009quasinormal,oliveira2019quasinormal,nollert1999quasinormal,horowitz2000quasinormal,kokkotas1999quasi,ferrari1984new,roy2020revisiting,maggiore2008physical,flachi2013quasinormal,ovgun2018quasinormal,london2014modeling,blazquez2018scalar,konoplya2011quasinormal} but also within the context of alternative gravity theories, which includes Ricci--based theories \cite{kim2018quasi, jawad2020quasinormal,lee2020quasi}, Lorentz violation \cite{maluf2013matter, maluf2014einstein}, and other related fields \cite{JCAP2,JCAP1,jcap4,JCAP3,jcap5, aa2023analysis}.

Significant progress has been made in the field of gravitational wave detection, enabling the identification of waves emitted by various physical phenomena \cite{abbott2017multi,abbott2016ligo,abbott2017gw170817,abbott2017gravitational}. Ground--based interferometers, including VIRGO, LIGO, TAMA--300, and EO--600, have played a pivotal role in these detections \cite{coccia1995gravitational,luck1997geo600,abramovici1992ligo,fafone2015advanced}. Over time, these detectors have substantially improved their precision, approaching a level of genuine sensitivity \cite{evans2014gravitational}. These aspects gained from these advancements have provided valuable knowledge about the nature of astrophysical entities, encompassing boson and neutron stars for instance.

The detection of gravitational waves has far--reaching consequences for the study of black holes. By observing the emitted gravitational radiation, it becomes possible to directly confirm the existence of perturbed black holes \cite{thorne2000probing}. The pioneering work of Regge and Wheeler focused on exploring the stability of Schwarzschild black holes, laying the foundation for investigations into black hole perturbations \cite{regge1957stability}. Subsequently, Zerilli made seminal contributions to the study of perturbations, significantly advancing our understanding in this field \cite{zerilli1974perturbation,zerilli1970effective}.

In recent years, there has been a significant surge of interest in the study of gravitational solutions involving scalar fields due to their remarkable characteristics. Notably, the behavior of black holes with nontrivial scalar fields has challenged the well--established ``no--hair theorem" \cite{herdeiro2015asymptotically}. This departure from the expected behavior has opened up exciting avenues of research, delving into the existence of long--lived scalar field patterns \cite{ayon2016analytic}, the exploration of exotic astrophysical scenarios such as gravastars \cite{pani2009gravitational,chirenti2016did,visser2004stable}, and the formation of boson stars \cite{palenzuela2017gravitational,colpi1986boson,cunha2017lensing}. Moreover, considering Klein--Gordon scalar fields on curved backgrounds has revealed a plethora of fascinating phenomena, including the intriguing concept of black hole bombs \cite{sanchis2016explosion,cardoso2004black,hod2016charged}, where the scalar field can enhance the extraction of energy from the black hole. Additionally, the phenomenon of superradiance \cite{brito2015black} arises when scalar fields interact with rotating black holes, leading to amplification effects.

In Verlinde's theory \cite{verlinde2011origin}, dark matter is proposed to emerge as a consequence of gravity arising from the distribution of baryonic matter. According to this theory, an additional gravitational effect is postulated due to the volume law contribution to entropy associated with positive dark energy. The hypothesis posits that the distribution of baryonic matter decreases the universe's overall entropy, triggering an elastic reaction in the fundamental microscopic system. This reaction gives rise to an extra gravitational force, commonly labeled as the dark matter effect, which is inherent to gravity's essence. Verlinde's theory offers a comprehensive model to interpret numerous observational phenomena, such as the rotation curves of galaxies. It proposes that the interplay between the distribution of baryonic matter and the perceived dark matter can elucidate the consistently observed flat rotation curves in galaxies.

In this study, our focus lies in examining a regular black hole within the framework of Verlinde's emergent gravity. Specifically, we place emphasis on investigating the modified Simpson--Visser solution. Through our analysis, we reveal the presence of a single physical event horizon. By studying the \textit{Hawking} temperature and heat capacity, we unveil the existence of phase transitions. Furthermore, we provide the calculation of geodesic trajectories for photon--like particles, including critical orbits known as photon spheres. Additionally, our study includes calculations for both time delay and deflection angle. Moreover, to substantiate our findings, we introduce an additional application within the realm of high--energy astrophysical phenomena: the deposition of neutrino energy.

investigate the quasinormal modes employing third--order WKB approximations.

\section{Regular black hole in Verlinde's gravity}

In our manuscript, our main focus is to correlate baryonic matter and apparent dark matter in order to develop black hole solutions within the framework of a theory VEG (Verlinde's emergent gravity) \cite{jusufi2023regular}. In other words, our goal is to investigate the effects arising from the existence of apparent dark matter on the spacetime geometry in the subsequent sections. 

Verlinde says that in the context of spherical symmetry, a connection exists between the quantities of apparent dark matter $M_D(r)$ and baryonic matter $M_B(r)$ \cite{jusufi2023regular}. This relationship can be mathematically written as follows:
\ie
\int^{r}_{0} \frac{M^{2}_{D}(\tilde{r})}{\tilde{r}^{2}}\mathrm{d}\tilde{r} = \frac{a_{0}M_{B}(r)r}{6}.
\fe
In this context, we have introduced a constant denoted by $a_0$. In our investigation, we focus on the simplest scenario involving spherically symmetric black hole solutions. These ones are described by the line element:
\ie
\mathrm{ds}^{2} = g_{\mu\nu}\mathrm{d}x^{\mu}\mathrm{d}x^{\nu}= g_{00}\mathrm{d}t^{2} + g_{11}\mathrm{d}r^{2} + g_{22}\mathrm{d}\theta^{2} + g_{33}\mathrm{d}\phi^{2},
\fe
where $f(r)=-g_{00}=g^{-1}_{11}= 1 - 2m(r)/r$, $g_{22}=r^{2}$, $g_{33}=r^{2}\sin^{2}\theta$, and $m(r)$, being given by \cite{verlinde2017emergent}
\ie
m(r) = 4\pi \int^{r}_{0} \left[ \rho_{B}(\tilde{r}) + \rho_{D}(\tilde{r}) \right]\tilde{r}^{2} \mathrm{d}\tilde{r}.
\fe
Building upon the recent findings presented in Ref. \cite{jusufi2023regular}, we aim to derive a regular black hole solution. Subsequently, we will explore the impact of dark matter on the geometry of the spacetime in the upcoming sections. It is important to highlight that in Ref. \cite{cadoni2022effective} the authors revisited the general properties of regular models with a dS core in light of Verlinde's idea. Additionally, we shall analyze the discernible features arising from such a spacetime, which include the event horizon, thermodynamics, geodesics, shadows, and the characteristics of \textit{quasinormal} modes.


\section{Thermodynamics}

We proceed by following the approach presented in reference \cite{simpson2019black}, where the metric describing the geometry of the spacetime is given by:
\ie
\mathrm{ds}^{2} = g_{00}\mathrm{d}t^{2} + g_{11}\mathrm{d}r^{2} + (r^{2} + a^{2})(\mathrm{d}\theta^{2} + \sin^{2}\theta \mathrm{d}\phi^{2} ),
\fe
with
\ie
M_{B}(r) = \frac{M r}{\sqrt{r^{2}+a^{2}}} - \frac{Q^{2}r}{2(r^{2}+a^{2})}.
\fe
If we consider that $Q \rightarrow 0$, the following solution gives rise to \cite{jusufi2023regular}:
\ie
\label{frrr}
f(r) = 1 - \frac{2M}{\sqrt{r^{2}+a^{2}}} - 2\sqrt{M}\sqrt{\frac{r(r^{2}+2a^{2})}{(r^{2}+a^{2})^{3/2}}}.
\fe
Above solution represents a generalization of the black--bounce spacetime geometry examined in reference \cite{simpson2019black}. However, a notable distinction arises from the presence of apparent dark matter, which renders the spacetime non--asymptotically flat.

Despite the presence of six roots in the aforementioned expression, it is important to note that only one of these roots corresponds to a physical horizon (labeled as $r_{+}$):
\ie
\begin{split}
r_{+} &= -\frac{1}{2} \sqrt{\frac{2304 a^4 M^4}{\sigma ^2}+\frac{\sqrt[3]{2} \kappa }{3 \gamma  \sigma }+\frac{\gamma }{3 \sqrt[3]{2} \sigma }-\frac{2 \eta }{3 \sigma }}-\frac{24 a^2 M^2}{\sigma } + \mathcal{O}(M^{5})+ \mathcal{O}(M^{6}) \\
& + \frac{1}{2} \sqrt{\frac{4608 a^4 M^4}{\sigma ^2}-\frac{-\frac{884736 a^6 M^6}{\sigma ^3}-\frac{512 a^4 M^2}{\sigma ^2}+\frac{384 a^2 \eta  M^2}{\sigma ^2}}{4 \sqrt{\frac{2304 a^4 M^4}{\sigma ^2}+\frac{\sqrt[3]{2} \kappa }{3 \gamma  \sigma }+\frac{\gamma }{3 \sqrt[3]{2} \sigma }-\frac{2 \eta }{3 \sigma }}}-\frac{\sqrt[3]{2} \kappa }{3 \gamma  \sigma }-\frac{\gamma }{3 \sqrt[3]{2} \sigma }-\frac{4 \eta }{3 \sigma }},
\end{split}
\fe
where, 
\ie
\sigma =64 a^2 M^2-3 a^2+8 M^2,
\fe
\ie
\eta =64 a^4 M^2-3 a^4+16 a^2 M^2-16 M^4,
\fe
\ie
\begin{split}
\kappa = & 4096 a^8 M^4-1152 a^8 M^2+45 a^8-10240 a^6 M^4-480 a^6 M^2 \\
& - 14336 a^4 M^6+1696 a^4 M^4-2048 a^2 M^6+256 M^8,
\end{split}
\fe
\ie
\begin{split}
\gamma = & \left[ 110592 a^8 M^4 \left(64 a^2 M^2-3 a^2+8 M^2\right)-55296 a^6 M^4 \left(64 a^4 M^2-3 a^4+16 a^2 M^2-16 M^4\right) \right.\\
& \left.  + 2 \left(64 a^4 M^2-3 a^4+16 a^2 M^2-16 M^4\right)^3+248832 a^4 M^4 \left(-a^6+8 a^4 M^2-16 a^2 M^4\right) \right.\\
& \left.  + 72 \left(64 a^2 M^2-3 a^2+8 M^2\right) \left(64 a^4 M^2-3 a^4+16 a^2 M^2-16 M^4\right) \left(-a^6+8 a^4 M^2-16 a^2 M^4\right) \right.\\
& \left.  -4 \left(4096 a^8 M^4-1152 a^8 M^2+45 a^8-10240 a^6 M^4-480 a^6 M^2-14336 a^4 M^6 \right.\right. \\
 & \left.\left. + 1696 a^4 M^4-2048 a^2 M^6+256 M^8\right)^3 + \left(110592 a^8 M^4 \left(64 a^2 M^2-3 a^2+8 M^2\right) \right.\right.\\
 & \left.\left. - 55296 a^6 M^4 \left(64 a^4 M^2-3 a^4+16 a^2 M^2-16 M^4\right) \right.\right.\\
 & \left.\left. +2 \left(64 a^4 M^2-3 a^4+16 a^2 M^2-16 M^4\right)^3+248832 a^4 M^4 \left(-a^6+8 a^4 M^2-16 a^2 M^4\right) \right.\right.\\
 & \left.\left. - 72 \left(64 a^2 M^2-3 a^2+8 M^2\right)\left(64 a^4 M^2-3 a^4+16 a^2 M^2-16 M^4\right) \left(-a^6+8 a^4 M^2-16 a^2 M^4\right) \right)^{2}\right]^{3/2},
\end{split}
\fe
and $\mathcal{O}(M^{5})$ and $\mathcal{O}(M^{6})$ are higher--order mass terms, i.e., fifth-- and sixth-- orders respectively. It is important to mention that all terms (even the higher order ones) will be taken into account in our calculations.

\begin{figure}
    \centering
    \includegraphics[scale=0.51]{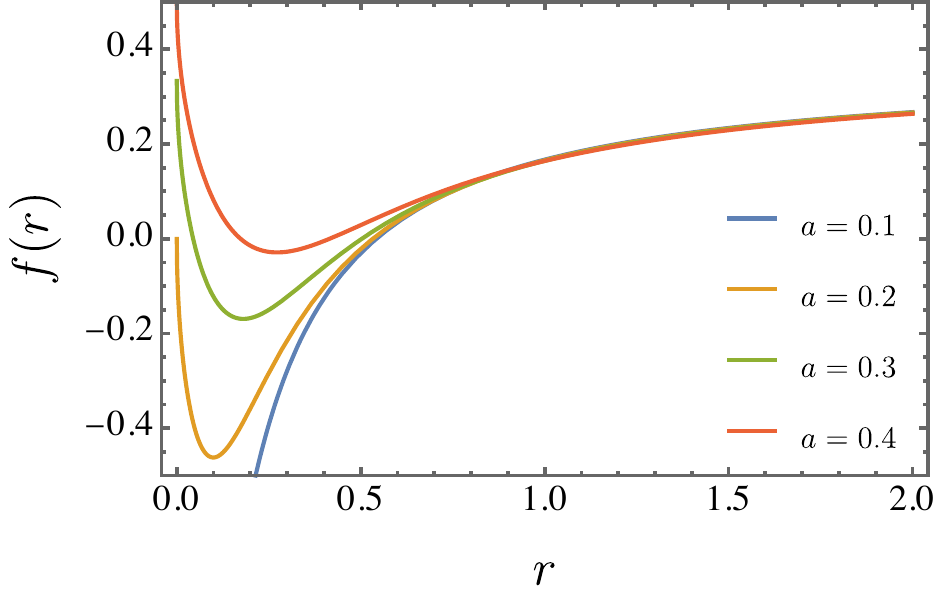}
    \includegraphics[scale=0.51]{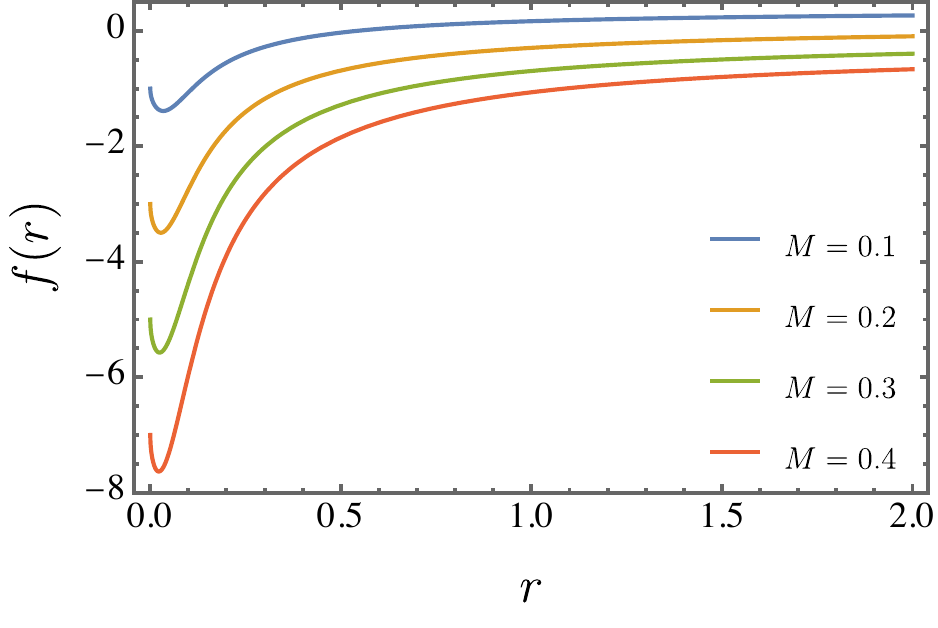}
    \caption{On the left side, we depict $f(r)$ for various values of $a$ (with $M=0.1$), while on the right side, we illustrate $f(r)$ for different values of $M$ (with $a=0.1$).}
    \label{frsss}
\end{figure}

\begin{figure}
    \centering
    \includegraphics[scale=0.4]{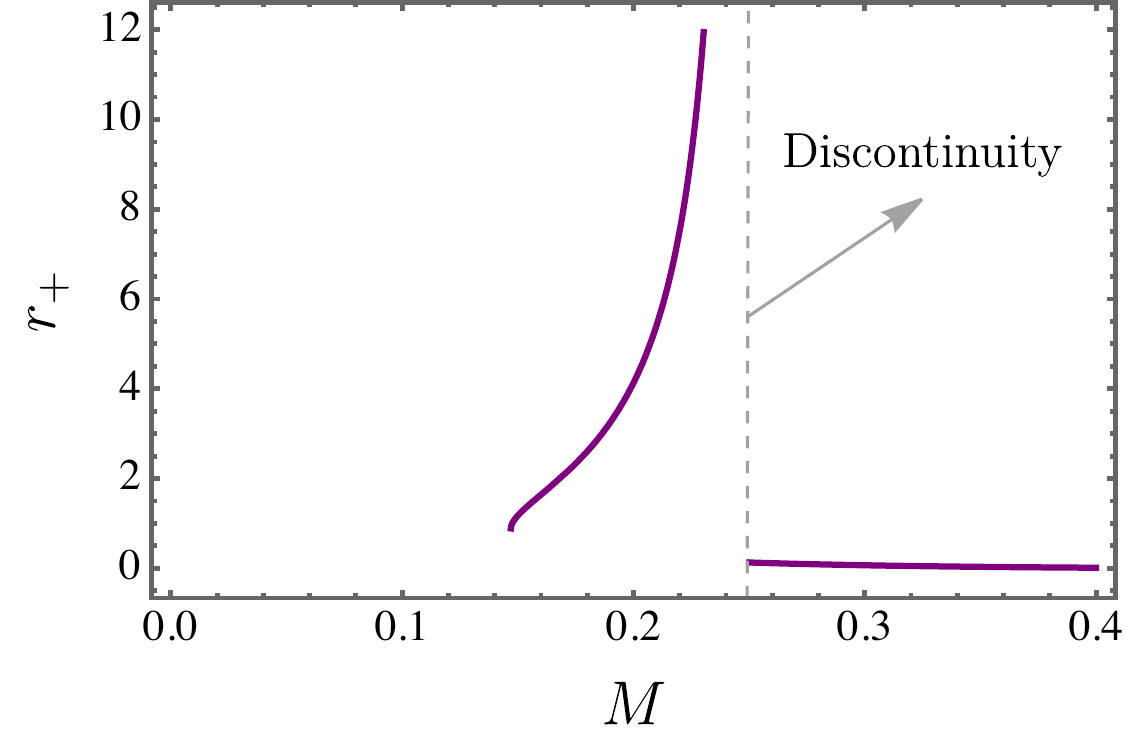}
    \includegraphics[scale=0.4]{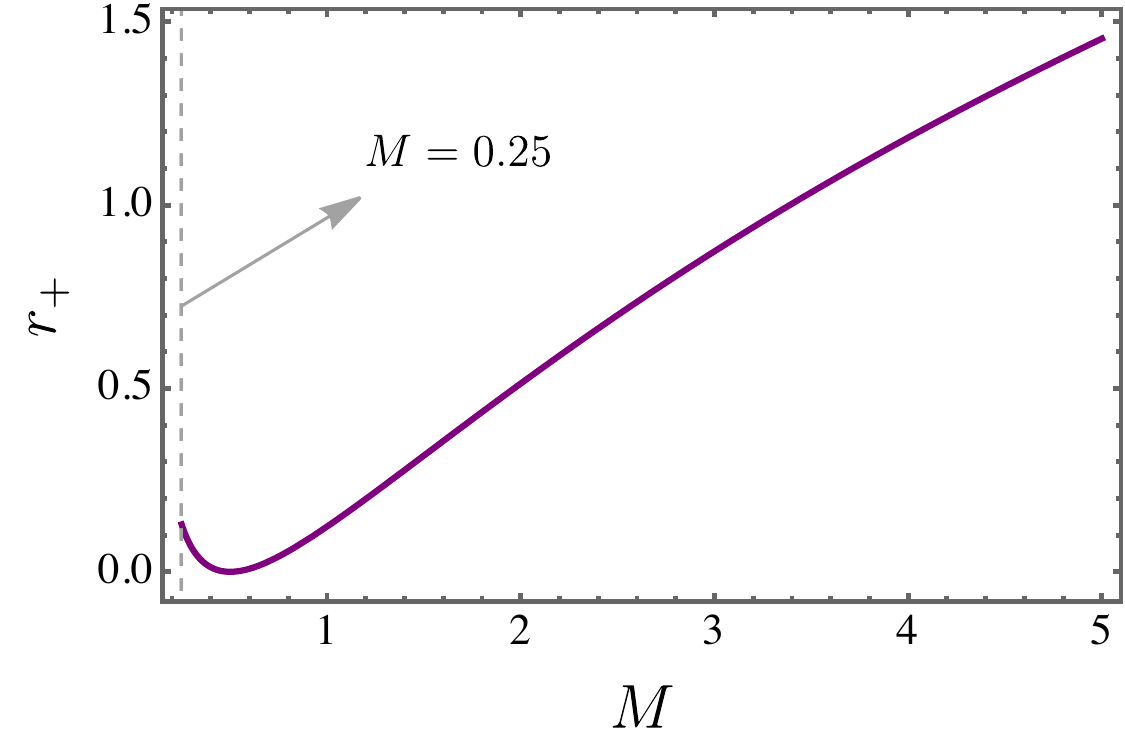}
    \includegraphics[scale=0.4]{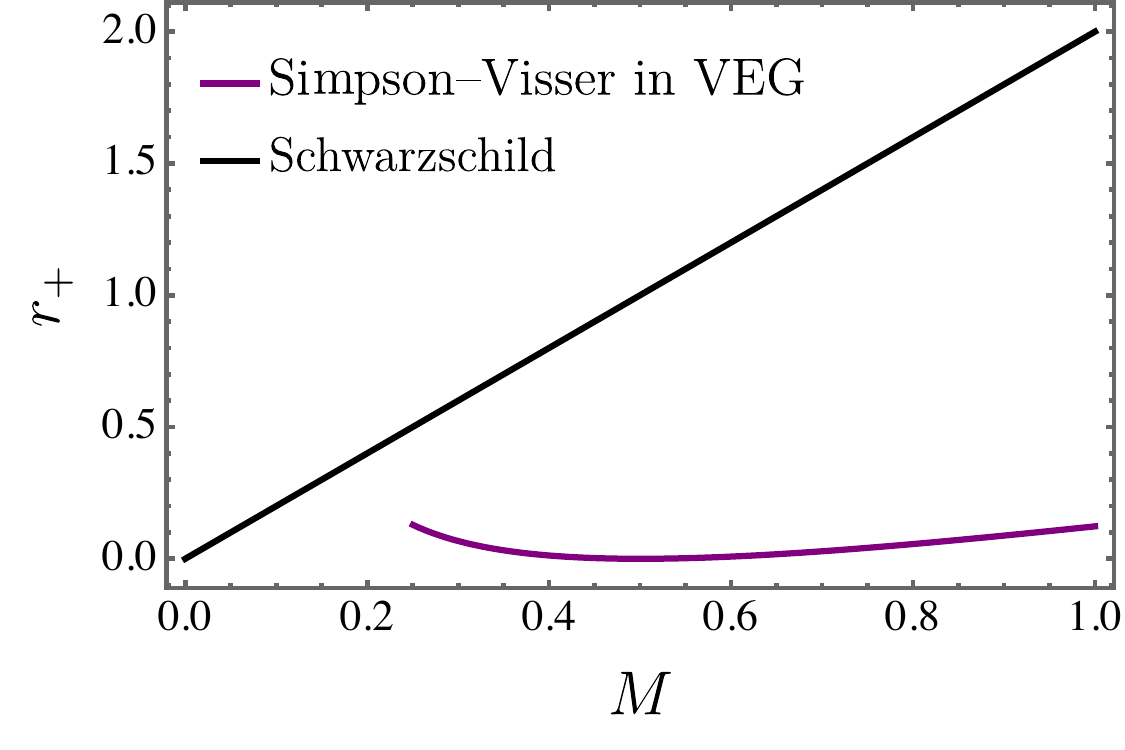}
    \caption{The modified Simpson--Visser horizon, $r_{+}$, is displayed. A discontinuity occurs when $M=0.25$ and  the mass must obey the following constraint: $M > 0.14699$ (when $a = 1$). It is also shown the comparison between the Simpson--Visser horizon and the Schwarzschild one (the bottom plot).}
    \label{eventhorizon}
\end{figure}

In order to get a better comprehension of Eq. (\ref{frrr}), we provide Fig. \ref{frsss}. To the left, we showcase $f(r)$ across different values of $a$ with $M=0.1$, and to the right, we present $f(r)$ for varying $M$ with $a=0.1$. Here, the horizon $r_{+}$ has a particularity, i.e., for possessing real positive defined values, the mass should should satisfy the following constraint: $M > 0.14699$ (when $a = 1$). Notice that such a condition will limit the whole thermodynamic properties of system from below. More so, a point of singularity is encountered here when $M=0.25$ (for $a=1$). All these features as well as the comparison with the Schwarzschild case are displayed in Fig. \ref{eventhorizon}.

Upon obtaining the event horizon, we can now accurately examine the modified \textit{Hawking} temperature as follows
\ie
\begin{split}
T = &-\frac{1}{4\pi} \frac{1}{\sqrt {-{g_{00}}{g_{11}}} }{\left. {\frac{{\mathrm{d}{g_{00}}}}{{\mathrm{d}r}}} \right|_{r = {r_{+}}}}\\
 =& - \frac{1}{4\pi} \left[-\frac{2 M r_{+}}{\left(r_{+}^2+1\right)^{3/2}}-\frac{3 \sqrt{M} \sqrt{r_{+}^2+1} \sqrt{r_{+} \left(r_{+}^2+2\right)} r_{+}}{\left(\left(r_{+}^2+1\right)^{3/2}\right)^{3/2}}+\frac{\sqrt{M} \left(3 r_{+}^2+2\right)}{\sqrt{\left(r_{+}^2+1\right)^{3/2}} \sqrt{r_{+} \left(r_{+}^2+2\right)}} \right].
 \label{temperaturesurfacegravity}
\end{split}
\fe
\begin{figure}
    \centering
    \includegraphics[scale=0.335]{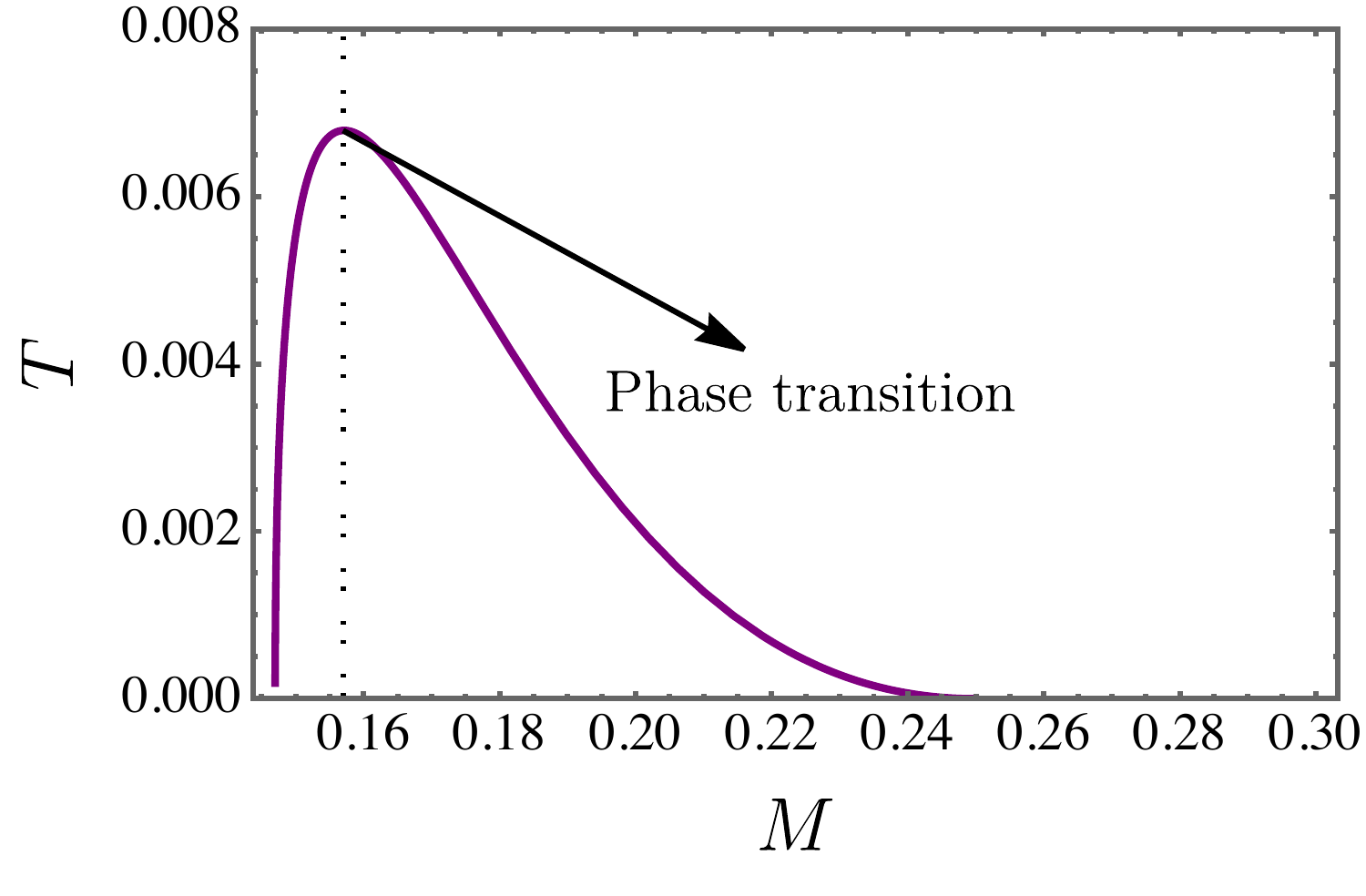}
    \caption{The modified \textit{Hawking} temperature is displayed, emphasizing the occurrence of a phase transition at $M \approx 0.1568$.}
    \label{temperature}
\end{figure}
Figure \ref{temperature} illustrates the behavior of the modified \textit{Hawking} temperature as a function of mass. Remarkably, our results suggest the occurrence of a phase transition when $M \approx 0.1568$. These remarks can be directly attributed to the influence of dark matter effects on the spherically symmetric black hole under investigation. This observation highlights the significant role of dark matter in shaping the thermodynamic properties of the black hole system.

Now, let us use the first law of thermodynamics to calculate the \textit{Hawking} temperature as well. Thereby, we can write the mass as
\ie
\begin{split}
M_{\pm} = &\frac{1}{8} \left[\left(\frac{4 r_{+}^3}{a^2+r_{+}^2}-8 r_{+}-4\right) \left(-\sqrt{a^2+r_{+}^2}\right) \right. \\
& \left. \pm \sqrt{\left(a^2+r_{+}^2\right) \left(\frac{4 r_{+}^3}{a^2+r_{+}^2}-8 r_{+}-4\right)^2-16 \left(a^2+r_{+}^2\right)}\right].
\end{split}
\fe
For our purpose, we shall consider $M_{+}$ only. Among other reasons, when we consider $a=1$ and $r$ runs, the unique solution which gives rise to a real positive defined values of mass is $M_{+}$. In this sense, we can perform the calculation of the \textit{Hawking} temperature via first law of thermodynamics, $\Tilde{T}$, as follows
\ie
\begin{split}
\Tilde{T} = & \frac{\mathrm{d}M_{+}}{\mathrm{d}S}= \frac{1}{2\pi r_{+}} \frac{\mathrm{d}M_{+}}{\mathrm{d}r_{+}} \\
& = \frac{1}{16 \pi r_{+}} \left[  \left(\frac{12 r_{+}^2}{a^2+r_{+}^2}-\frac{8 r_{+}^4}{\left(a^2+r_{+}^2\right)^2}-8\right) \left(-\sqrt{a^2+r_{+}^2}\right)-\frac{r_{+} \left(\frac{4 r_{+}^3}{a^2+r_{+}^2}-8 r_{+}-4\right)}{\sqrt{a^2+r_{+}^2}} \right. \\
& \left.  +\frac{2 r_{+} \left(\frac{4 r_{+}^3}{a^2+r_{+}^2}-8 r_{+}-4\right)^2+2 \left(a^2+r_{+}^2\right) \left(\frac{12 r_{+}^2}{a^2+r_{+}^2}-\frac{8 r_{+}^4}{\left(a^2+r_{+}^2\right)^2}-8\right) \left(\frac{4 r_{+}^3}{a^2+r_{+}^2}-8 r_{+}-4\right)-32 r_{+}}{2 \sqrt{\left(a^2+r_{+}^2\right) \left(\frac{4 r_{+}^3}{a^2+r_{+}^2}-8 r_{+}-4\right)^2-16 \left(a^2+r_{+}^2\right)}}    \right].
\label{temperaturefirstlaw}
\end{split}
\fe

Notice that Eqs. (\ref{temperaturefirstlaw}) and (\ref{temperaturesurfacegravity}) did not match to each other. In other words, it is clearly seen that the first law of thermodynamics did not give us the correct expression. In general lines, this is because regular black holes do not obey the so--called \textit{Bekenstein--Hawking} area law.
For the sake of overcoming this situation,  a correction must be implemented in Eq. (\ref{temperaturefirstlaw}) \cite{ma2014corrected}. Instead, the corrected temperature is written as \cite{ma2014corrected}
\ie
\Upsilon(r_{+},a) \mathrm{d}M = \overset{\nsim}{T}\, \mathrm{d}S.
\fe
In this context, $\overset{\nsim}{T}$ represents the corrected version of the \textit{Hawking} temperature obtained through the application of the first law of thermodynamics, with $S$ denoting the entropy. Notably, the function $\Upsilon(r_{+}, a)$, intricately dependent on the terms of the mass function, not only governs the first law for regular black holes within the specific scenario under consideration but also assumes a crucial role in formulating the first law for diverse classes of regular black holes, as argued in \cite{maluf2018thermodynamics}.

As highlighted in Ref. \cite{ma2014corrected}, the general expression for $\Upsilon(r_{+},a)$ is articulated as follows:
\ie
\Upsilon(r_{+},a) = 1 + 4\pi \int^{\infty}_{r_{+}} r^{2} \frac{\partial T^{0}_{0}}{\partial  M_{+}} \mathrm{d}r.
\fe
Here, the notation $T^{0}_{0}$ refers to the stress--energy component corresponding to energy density. Specifically, it is defined as
\ie
\begin{split}
T^{0}_{0} = & -\frac{\sqrt{M}}{8\pi r^2 \left(a^2+r^2\right)^2 \sqrt{2 a^2 r+r^3} \Box^2} \\
& \times \left[ 24 M^2 r^7 (2 r-1) \sqrt{\left(a^2+r^2\right)^{3/2}}-2 r^9 \sqrt{\left(a^2+r^2\right)^{3/2}} \right. \\
& \left. +16 M^3 r^5 \sqrt{a^2+r^2} \sqrt{\left(a^2+r^2\right)^{3/2}}-12 M r^7 (2 r-1) \sqrt{a^2+r^2} \sqrt{\left(a^2+r^2\right)^{3/2}} \right. \\
& \left. +48 M^{5/2} r^7 \sqrt{2 a^2 r+r^3}+12 \sqrt{M} r^9 \sqrt{2 a^2 r+r^3} \right. \\
& \left. +16 M^{3/2} (r-3) r^7 \sqrt{a^2+r^2} \sqrt{2 a^2 r+r^3} \right. \\ 
& \left. -2 a^8 \left(\sqrt{M} (1-18 r) \sqrt{2 a^2 r+r^3}+3 r \sqrt{\left(a^2+r^2\right)^{3/2}}\right) \right. \\
& \left. +a^2 r^2\left(4 M^2 r^3 (58 r-33) \sqrt{\left(a^2+r^2\right)^{3/2}}-9 r^5 \sqrt{\left(a^2+r^2\right)^{3/2}}\right. \right. \\
& \left. +88 M^3 r \sqrt{a^2+r^2} \sqrt{\left(a^2+r^2\right)^{3/2}}-54 M r^3 (2 r-1) \sqrt{a^2+r^2} \sqrt{\left(a^2+r^2\right)^{3/2}} \right. \\
& \left. +24 M^{5/2} r^2 (9 r-1) \sqrt{2 a^2 r+r^3}+2 \sqrt{M} r^4 (27 r-1) \sqrt{2 a^2 r+r^3}  \right. \\
& \left. \left.12 M^{3/2} r^2 \left(6 r^2-16 r+1\right) \sqrt{a^2+r^2} \sqrt{2 a^2 r+r^3}+16 M^{7/2} \sqrt{a^2+r^2} \sqrt{2 a^2 r+r^3}\right) \right. \\
& \left. +a^6\left(8 M^2 r (44 r-21) \sqrt{\left(a^2+r^2\right)^{3/2}}-17 r^3 \sqrt{\left(a^2+r^2\right)^{3/2}}\right. \right. \\
& \left.  24 M^{5/2} (10 r-1) \sqrt{2 a^2 r+r^3}-12 M r (12 r-5) \sqrt{a^2+r^2} \sqrt{\left(a^2+r^2\right)^{3/2}} \right. \\
&\left.  \left.12 M^{3/2} \left(8 r^2-16 r+1\right) \sqrt{a^2+r^2} \sqrt{2 a^2 r+r^3}+6 \sqrt{M} r^2 (17 r-1) \sqrt{2 a^2 r+r^3}\right) \right. \\
& \left. +2 a^4 \left(2 M^2 r^3 (112 r-69) \sqrt{\left(a^2+r^2\right)^{3/2}}-9 r^5 \sqrt{\left(a^2+r^2\right)^{3/2}}\right.  \right. \\
& \left.  +72 M^3 r \sqrt{a^2+r^2} \sqrt{\left(a^2+r^2\right)^{3/2}}-3 M r^3 (32 r-17) \sqrt{a^2+r^2} \sqrt{\left(a^2+r^2\right)^{3/2}} \right. \\
&\left.  +12 M^{5/2} r^2 (17 r-2) \sqrt{2 a^2 r+r^3}+3 \sqrt{M} r^4 (18 r-1) \sqrt{2 a^2 r+r^3} \right.\\
& \left.   4 M^{3/2} r^2 \left(16 r^2-42 r+3\right) \sqrt{a^2+r^2} \sqrt{2 a^2 r+r^3}+ 8 M^{7/2} \sqrt{a^2+r^2} \sqrt{2 a^2 r+r^3}    \right],
\end{split}
\fe
where $\Box = 2 M \sqrt{\left(a^2+r^2\right)^{3/2}}+2 \sqrt{M} \sqrt{a^2+r^2} \sqrt{2 a^2 r+r^3}-\sqrt{a^2+r^2} \sqrt{\left(a^2+r^2\right)^{3/2}}$.

With all these features, we can properly calculate $\Upsilon(r_{+},a)$ as follows
\ie
\begin{split}
& \Upsilon(r_{+},a) = 1 + 4\pi \int^{\infty}_{r_{+}} r^{2}  \left\{  \frac{1}{16 \pi  \sqrt{M_{+}} r^2 \left(a^2+r^2\right)^2 \sqrt{2 a^2 r+r^3} \Box}  \times \left[ \right.\right. \\
& \left.\left. -6 a^{12} r - 2 r^7 \left(   48 M_{+}^4 r^2+16 M_{+}^2 (2 (r-6) r+3) r^4+r^6  +80 M_{+}^3 r^2 (3 r-1) \sqrt{a^2+r^2} \right.\right.\right. \\
& \left.\left.\left.   + 176 M_{+}^{7/2} \sqrt{\left(a^2+r^2\right)^{3/2}} \sqrt{2 a^2 r+r^3}+12 M_{+} r^4 (3 r-1) \sqrt{a^2+r^2} \right.\right.\right. \\
& \left.\left.\left. -28 M_{+}^{3/2} r^2 (2 r-3) \sqrt{\left(a^2+r^2\right)^{3/2}} \sqrt{2 a^2 r+r^3}  - 10 \sqrt{M_{+}} r^2 \sqrt{a^2+r^2} \sqrt{\left(a^2+r^2\right)^{3/2}} \sqrt{2 a^2 r+r^3}  \right.\right.\right. \\
& \left.\left.\left. +72 M_{+}^{5/2} (2 r-3) \sqrt{a^2+r^2} \sqrt{\left(a^2+r^2\right)^{3/2}} \sqrt{2 a^2 r+r^3} \right)           - 2 a^8 \right.\right. \\
& \left.\left. \times \left(  24 M_{+} (30 r-11) r^3 \sqrt{a^2+r^2}+624 M_{+}^4 r+29 r^5   +32 M_{+}^3 r (71 r-27) \sqrt{a^2+r^2}  \right.\right.\right. \\
& \left.\left.\left. -2 \sqrt{M_{+}} (15 r-1) \sqrt{a^2+r^2} \sqrt{\left(a^2+r^2\right)^{3/2}} \sqrt{2 a^2 r+r^3} \right.\right.\right. \\
& \left.\left.\left. -4 M_{+}^{3/2} \left(84 r^2-93 r+5\right) \sqrt{\left(a^2+r^2\right)^{3/2}} \sqrt{2 a^2 r+r^3}   +4 M_{+}^2 r^3 (136 r (2 r-9)+369)            \right)  \right.\right. \\
& \left.\left. - a^{10} r \left(144 M_{+} (3 r-1) \sqrt{a^2+r^2}+16 M_{+}^2 (4 r (12 r-47)+51)+29 r^2\right)    \right.\right. \\
& \left.\left.   -a^2 r^2 \left(     12 M_{+} (39 r-14) r^7 \sqrt{a^2+r^2}+912 M_{+}^4 r^5+13 r^9  \right.\right.\right. \\
& \left.\left.\left.   +16 M_{+}^3 (191 r-84) r^5 \sqrt{a^2+r^2}+8 M_{+}^2 \left(60 r^2-334 r+99\right) r^7  \right.\right.\right. \\
& \left.\left.\left.  + 112 M_{+}^{7/2} r^2 (17 r-2) \sqrt{\left(a^2+r^2\right)^{3/2}} \sqrt{2 a^2 r+r^3} + 128 M_{+}^{9/2} \sqrt{a^2+r^2} \sqrt{\left(a^2+r^2\right)^{3/2}} \sqrt{2 a^2 r+r^3}  \right.\right.\right. \\
& \left.\left.\left.   - 2 \sqrt{M_{+}} r^4 (45 r-2) \sqrt{a^2+r^2} \sqrt{\left(a^2+r^2\right)^{3/2}} \sqrt{2 a^2 r+r^3} \right.\right.\right. \\
& \left.\left.\left. +24 M_{+}^{5/2} r^2 (29 r (2 r-3)+6) \sqrt{a^2+r^2} \sqrt{\left(a^2+r^2\right)^{3/2}} \sqrt{2 a^2 r+r^3}  \right.\right.\right. \\
& \left.\left.\left.  -4 M_{+}^{3/2} r^4 (r (154 r-219)+10) \sqrt{\left(a^2+r^2\right)^{3/2}} \sqrt{2 a^2 r+r^3}                   \right)  \right.\right. \\
& \left.\left. -2 a^4 \left(           8 M_{+}^3 (505 r-246) r^5 \sqrt{a^2+r^2}+36 M_{+} (18 r-7) r^7 \sqrt{a^2+r^2}+1392 M_{+}^4 r^5+19 r^9 \right.\right.\right. \\
& \left.\left.\left. + 8 M_{+}^{7/2} r^2 (247 r-28) \sqrt{\left(a^2+r^2\right)^{3/2}} \sqrt{2 a^2 r+r^3}+64 M_{+}^{9/2} \sqrt{a^2+r^2} \sqrt{\left(a^2+r^2\right)^{3/2}} \sqrt{2 a^2 r+r^3}   \right.\right.\right. \\
& \left.\left.\left.  - 6 \sqrt{M_{+}} r^4 (15 r-1) \sqrt{a^2+r^2} \sqrt{\left(a^2+r^2\right)^{3/2}} \sqrt{2 a^2 r+r^3} +4 M_{+}^2 r^7 (4 r (47 r-249)+327)    \right.\right.\right. \\
& \left.\left.\left.    +12 M_{+}^{5/2} r^2 (r (112 r-171)+12) \sqrt{a^2+r^2} \sqrt{\left(a^2+r^2\right)^{3/2}} \sqrt{2 a^2 r+r^3}  \right.\right.\right. \\
& \left.\left.\left.   -4 M_{+}^{3/2} r^4 (r (175 r-249)+15) \sqrt{\left(a^2+r^2\right)^{3/2}} \sqrt{2 a^2 r+r^3}  \right)   \right.\right. \\
& \left.\left.   -2 a^6 \left(  32 M_{+}^3 (157 r-70) r^3 \sqrt{a^2+r^2}+1608 M_{+}^4 r^3+31 r^7 +6 M_{+} r^5 (159 r-62) \sqrt{a^2+r^2} \right.\right.\right. \\
& \left.\left.\left.   +16 M_{+}^{7/2} (75 r-7) \sqrt{\left(a^2+r^2\right)^{3/2}} \sqrt{2 a^2 r+r^3}   \right.\right.\right. \\
& \left.\left.\left. + \sqrt{M_{+}} (6-85 r) r^2 \sqrt{a^2+r^2} \sqrt{\left(a^2+r^2\right)^{3/2}} \sqrt{2 a^2 r+r^3}  \right.\right.\right. \\
& \left.\left.\left. +24 M_{+}^{5/2} (r (44 r-51)+3) \sqrt{a^2+r^2} \sqrt{\left(a^2+r^2\right)^{3/2}} \sqrt{2 a^2 r+r^3}  \right.\right.\right. \\
& \left.\left.\left. + 4 M_{+}^2 r^5 (2 r (156 r-779)+507) -2 M_{+}^{3/2} r^2 (r (392 r-507)+30) \sqrt{\left(a^2+r^2\right)^{3/2}} \sqrt{2 a^2 r+r^3}        \right)        \right]  \right\},
\end{split}
\fe
which yields
\ie
\Upsilon(r_{+},a) = -\frac{4 \Delta r_{+}}{-\varphi  \sqrt{a^2+r_{+}^2}-\frac{r_{+} \epsilon }{\sqrt{a^2+r_{+}^2}}+\frac{\left(\epsilon ^2-16\right) \left(-8 a^4 \epsilon +a^2 r_{+} \left(-4 r_{+} \epsilon +\epsilon ^2-16\right)+r_{+}^3 \left(-4 r_{+} \epsilon +\epsilon ^2-16\right)\right)}{\left(\left(\epsilon ^2-16\right) \left(a^2+r_{+}^2\right)\right)^{3/2}}},
\fe
where 
\ie
\varphi = -\frac{4 \left(2 a^4+a^2 r_{+}^2+r_{+}^4\right)}{\left(a^2+r_{+}^2\right)^2}, \nonumber
\fe
\ie
\epsilon =\frac{4 r_{+}^3}{a^2+r_{+}^2}-8 r_{+}-4, \nonumber
\fe
and 
\ie
\begin{split}
\Delta = &-\frac{1}{2 \sqrt{2} \left(r_{+}^2+1\right)^2 \sqrt{\left(r_{+}^2+1\right)^{3/2}} \sqrt{r_{+} \left(r_{+}^2+2\right)}} \times \\
& 
\left\{  \sqrt{4 \sqrt{\frac{r_{+} \left(2 a^2+r_{+}^2\right) \left(2 a^2 (r_{+}+1)+r_{+}^2 (r_{+}+2)\right)}{a^2+r_{+}^2}}+\frac{4 \left(2 a^2 r_{+}+a^2+r_{+}^3+r_{+}^2\right)}{\sqrt{a^2+r_{+}^2}}}      \right\} \times \\
& \left[   \frac{\sqrt{r_{+}^2+1} \sqrt{\left(r_{+}^2+1\right)^{3/2}} \sqrt{r_{+} \left(r_{+}^2+2\right)} r_{+} \sqrt{4 \sqrt{\frac{r_{+} \left(2 a^2+r_{+}^2\right) \left(2 a^2 (r_{+}+1)+r_{+}^2 (r_{+}+2)\right)}{a^2+r_{+}^2}}+\frac{4 \left(2 a^2 r_{+}+a^2+r_{+}^3+r_{+}^2\right)}{\sqrt{a^2+r_{+}^2}}}}{\sqrt{2}} \right. \\
& \left. +r_{+}^4-r_{+}^2-2 \right].
\nonumber
\end{split}
\fe

Following careful consideration, it is observed that the \textit{Hawking} temperatures are mutually consistent, as articulated by the subsequent relations:
\begin{equation}
{\overset{\nsim}{T}} = T = \Upsilon(r_{+},a) \Tilde{T}.
\end{equation}
Consequently, the entropy $S$ can be expressed with precision as
\begin{equation}
S = \int \frac{\Upsilon(r_{+},a)}{\overset{\nsim}{T}} \mathrm{d} M_{+} = \pi (r_{+} + a)^{2} = \frac{A}{4}.
\end{equation}

\begin{figure}
    \centering
    \includegraphics[scale=0.4]{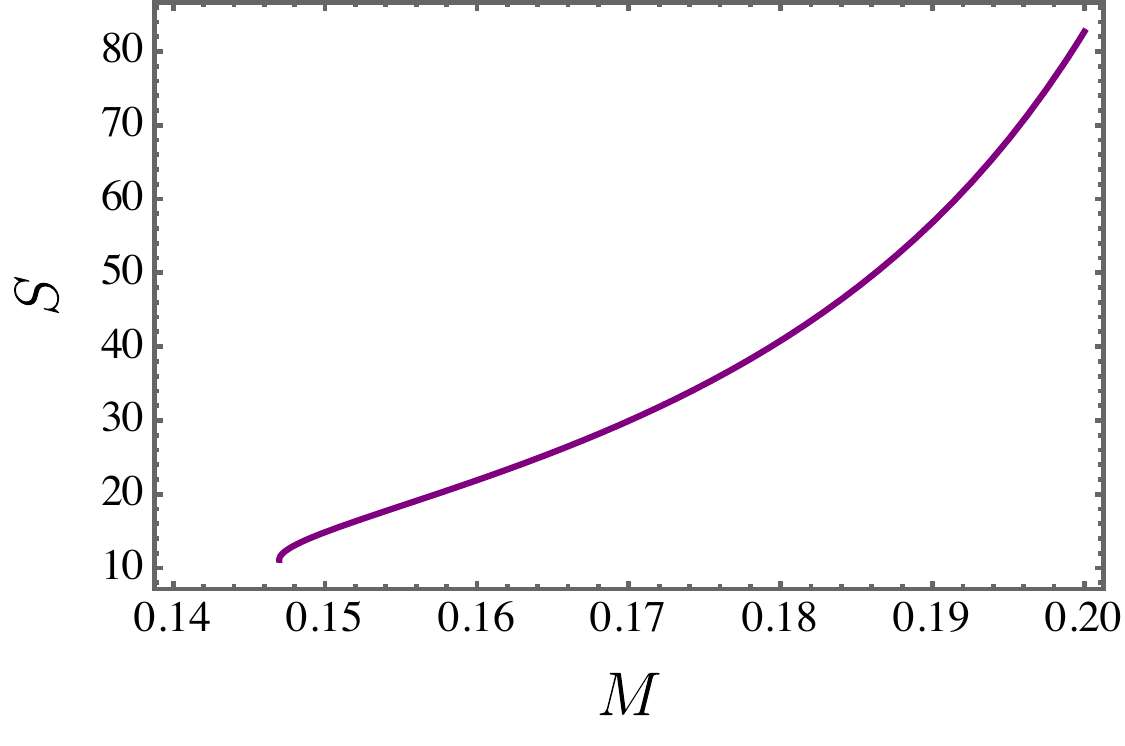}
    \includegraphics[scale=0.41]{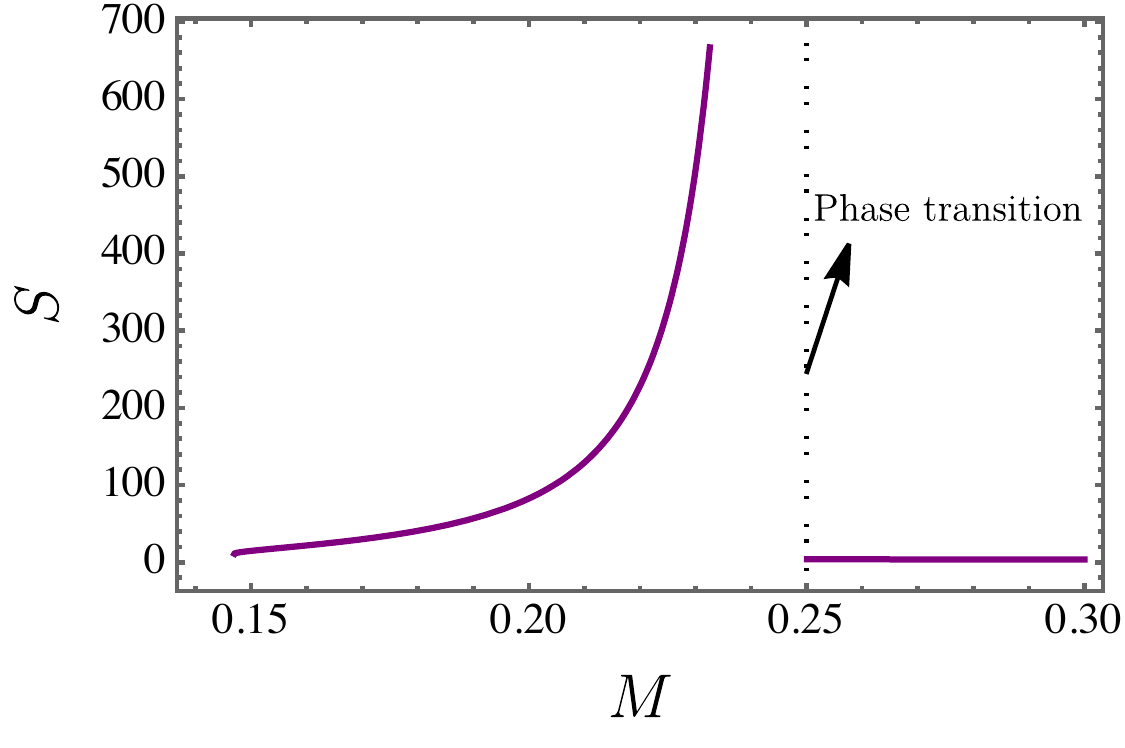}
    \includegraphics[scale=0.4]{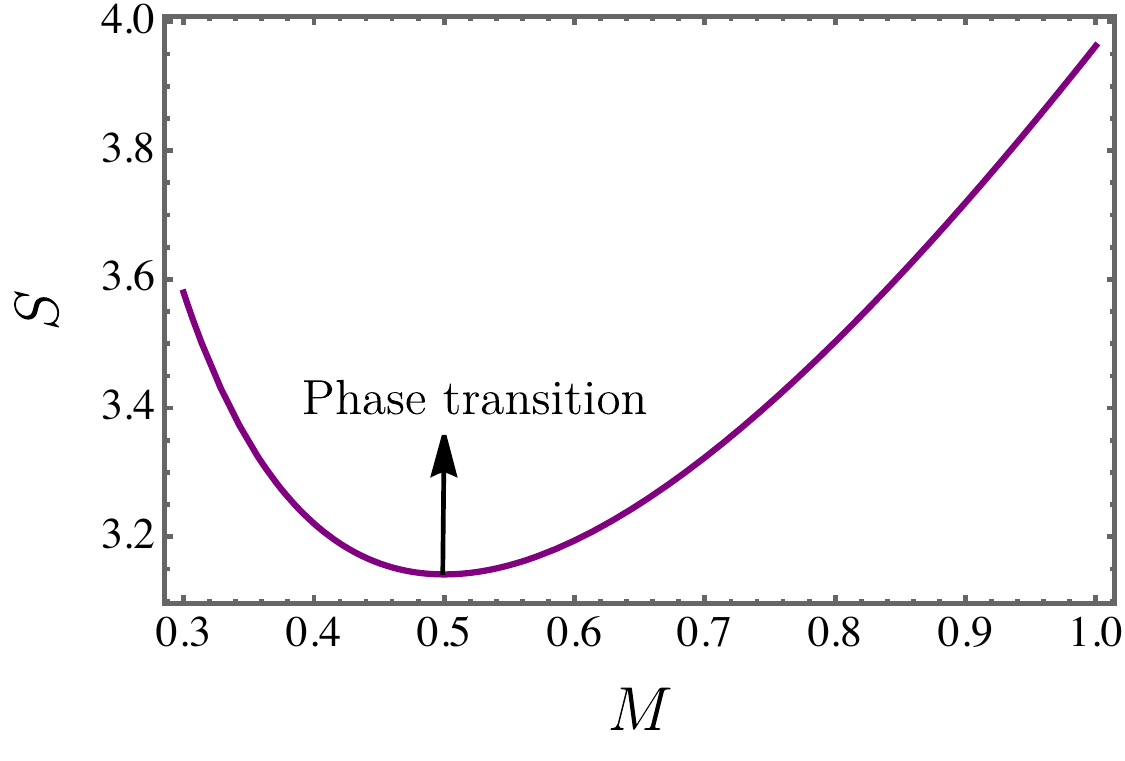}
    \caption{The modified entropy is displayed, highlighting multiple phase transition points.}
    \label{entropy}
\end{figure}

In order to get a better comprehension of such a thermodynamic function, we present Fig. \ref{entropy}. The entropy reveals at least two more points where the phase transition occurs, i.e., $M=0.25$ and $M=0.5$. Now, the last remaining thermodynamic quantity under consideration is the heat capacity

\ie
\begin{split}
& C_{V} = T \frac{\partial S}{\partial T} = T \frac{\partial S/\partial M}{\partial T/\partial M}\\
& = \frac{2 M \pi \left(-2 a^6-a^4 r_{+}^2+2 \sqrt{M} r_{+} \sqrt{a^2+r_{+}^2} \sqrt{\left(a^2+r_{+}^2\right)^{3/2}} \sqrt{2 a^2 r_{+}+r_{+}^3}+a^2 r_{+}^4\right)}{-2 a^6-a^4 r_{+}^2+4 \sqrt{M} r_{+} \sqrt{a^2+r_{+}^2} \sqrt{\left(a^2+r_{+}^2\right)^{3/2}} \sqrt{2 a^2 r_{+}+r_{+}^3}+a^2 r_{+}^4}\frac{\partial (r_{+}^{2}+a^{2})}{\partial M}.
\end{split}
\fe

\begin{figure}
    \centering
    \includegraphics[scale=0.4]{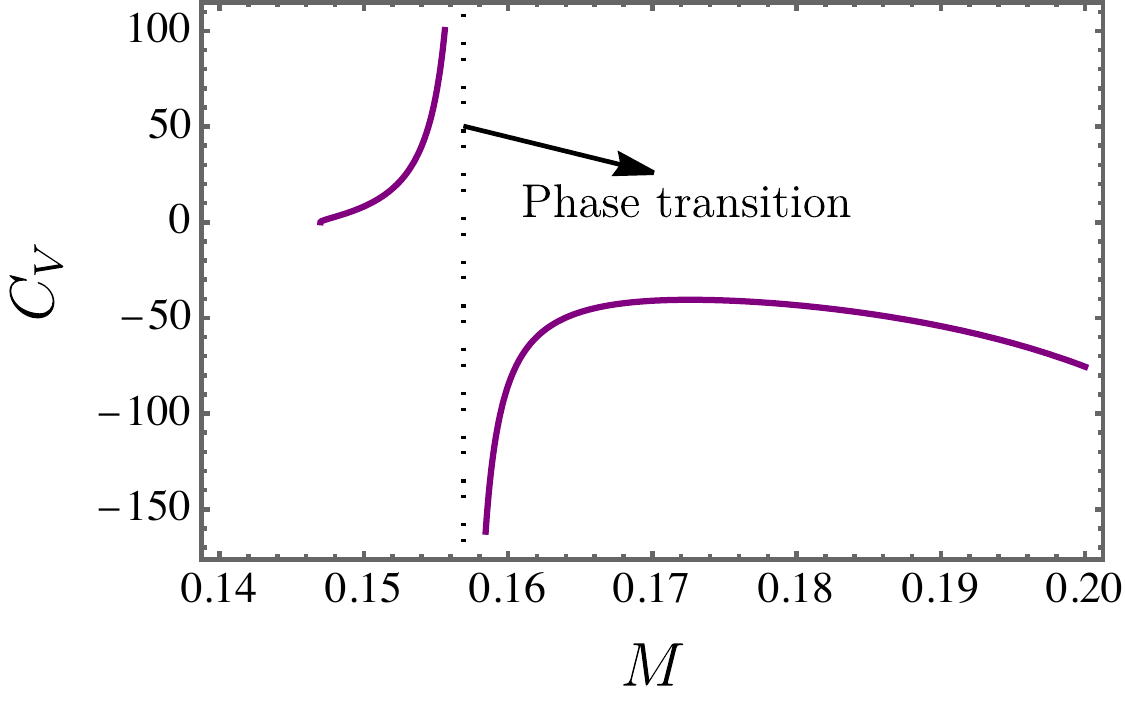}
    \includegraphics[scale=0.4]{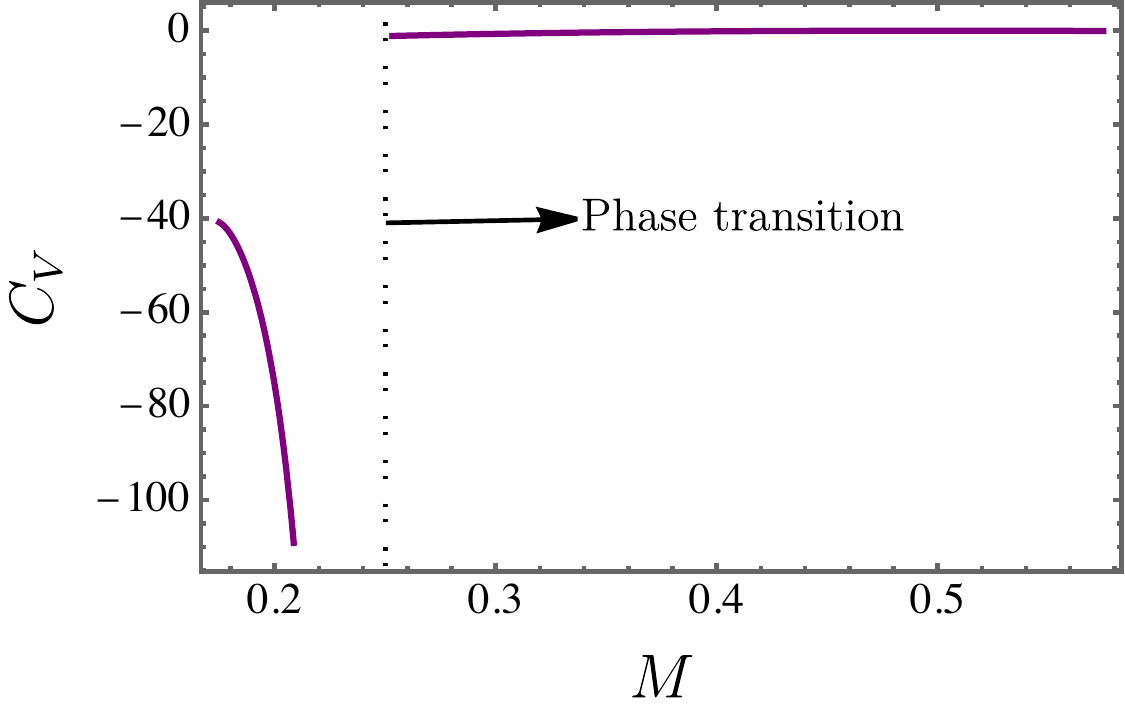}
     \includegraphics[scale=0.4]{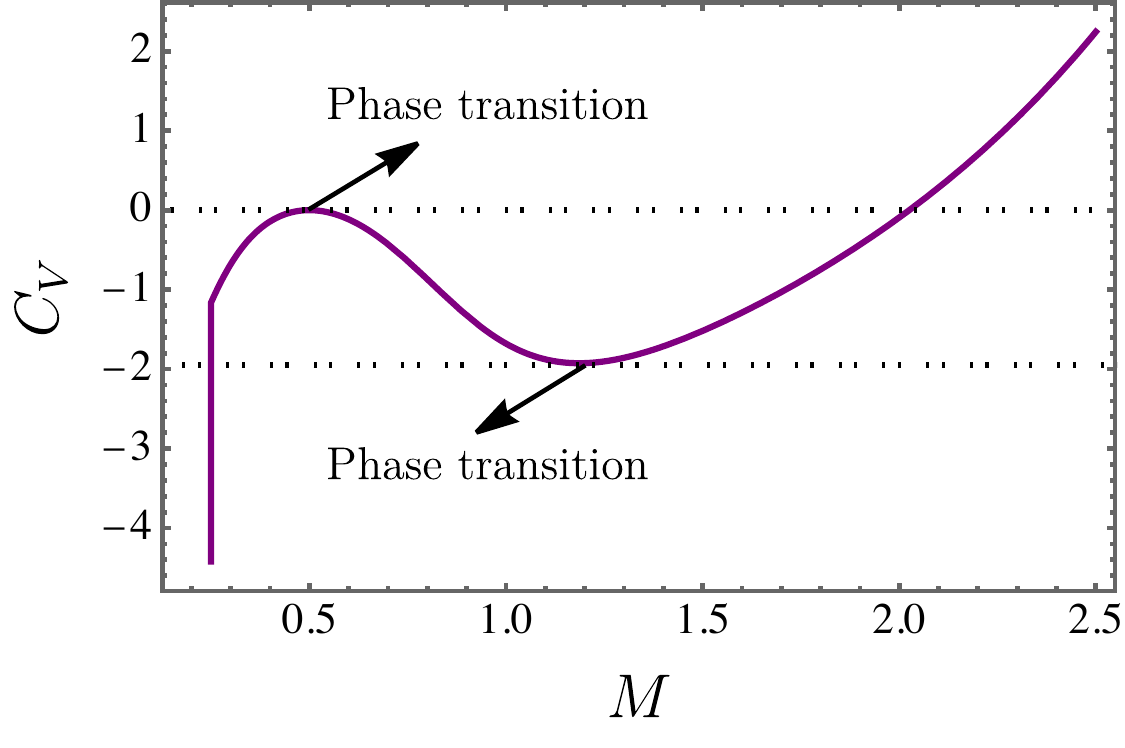}
    \caption{The modified heat capacity is displayed, highlighting multiple phase transition points.}
    \label{heatcapacity}
\end{figure}

Figure \ref{heatcapacity} depicts the heat capacity for different values of mass $M$. Consistent with our analysis of the \textit{Hawking} temperature, we observe a phase transition occurring around $M\approx 0.1568$. This feature also occurs for other configurations of the system, specifically $M=0.25$, $M=0.5$, and $M=1.2$. The first two values of mass are supported by the entropy behavior. Notably, a region of stability emerges, corresponding to the following range of heat capacity: $M \approx 0.14699$ to $M \approx 0.1568$.

These implications are only made possible by the presence of exotic matter, as it was introduced in the previous sections. Also, Table \ref{thermo} summarizes all thermodynamic state quantities. In addition, it is important to emphasize that the thermodynamic aspects have been extensively investigated across various contexts, including cosmological scenarios \cite{anacleto2021noncommutative,maluf2018thermodynamics,aa15,araujo2022thermal,araujo2023thermodynamics,aa2,araujo2021bouncing,anacleto2018lorentz,aa13,aa4,aa7,aa10,silva2013quantum,aa14,aa6,campos2022quasinormal,filho2022thermodynamics,hassanabadi2023fermions} and others \cite{aa11,qq,anacleto2018lorentz,aguirre2021lorentz,aa1,aa12}.

\begin{table}[t]
	\centering
		\caption{\label{thermo} The thermodynamic properties of the Schwarzschild and Simpson--Visser like black holes.}
	\setlength{\arrayrulewidth}{0.3mm}
	\setlength{\tabcolsep}{23pt}
	\renewcommand{\arraystretch}{1}
	\begin{tabular}{c c c}
		\hline \hline
		~ & Schwarzschild & Simpson--Visser in VEG  \\ \hline
	    \,\,\,$T$ & $1/8\pi M$ & $\frac{M r_{+}}{2 \pi  \left(a^2+r_{+}^2\right)^{3/2}}+\frac{a^2 \sqrt{M} \left(r_{+}^2-2 a^2\right)}{4 \pi  \left(a^2+r_{+}^2\right) \sqrt{\left(a^2+r_{+}^2\right)^{3/2}} \sqrt{2 a^2 r_{+}+r_{+}^3}}$ \\
		\,\,$A$ & $16 \pi M^{2} $ &$4\pi( r_{+}+a)^{2}$  \\ \
		$S$ &  $4 \pi M^{2}$ & $\pi (r_{+}+a)^{2}$ \\ 
		\,\,\,\,\,$C_{V}$ &  $- 8\pi {M^2}$ & $  \frac{2 M \pi \left(-2 a^6-a^4 r_{+}^2+2 \sqrt{M} r_{+} \sqrt{a^2+r_{+}^2} \sqrt{\left(a^2+r_{+}^2\right)^{3/2}} \sqrt{2 a^2 r_{+}+r_{+}^3}+a^2 r_{+}^4\right)}{-2 a^6-a^4 r_{+}^2+4 \sqrt{M} r_{+} \sqrt{a^2+r_{+}^2} \sqrt{\left(a^2+r_{+}^2\right)^{3/2}} \sqrt{2 a^2 r_{+}+r_{+}^3}+a^2 r_{+}^4}\frac{\partial (r_{+}^{2}+a^{2})}{\partial M}$ \\ \hline\hline
	\end{tabular}
\end{table}

\section{Geodesic trajectories}

\begin{figure}
    \centering
    \includegraphics[scale=0.75]{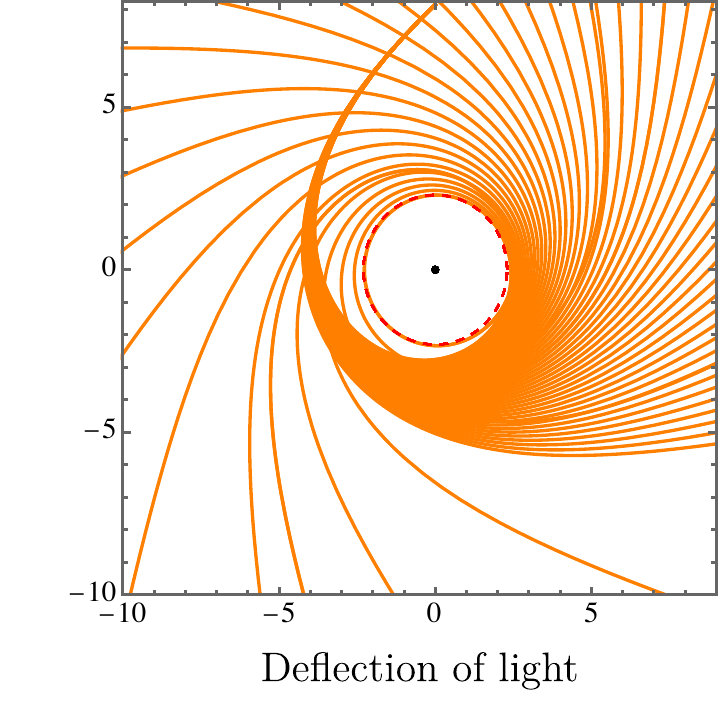}
    \caption{The depicted trajectories illustrate the paths of light, with the photon sphere indicated by red dashed lines. Additionally, the event horizon $r_{+}$ is denoted by a black dot. }
    \label{raytracing}
\end{figure}

The study of particle motion within the realm of emergent gravity has garnered significant interest owing to its profound implications from a theoretical viewpoint \cite{aa2023analysis,lim2018field,verlinde2011origin}. Of particular significance is comprehending the geodesic characteristics of Simpson--Visser black holes, as it holds crucial relevance in understanding various astrophysical phenomena associated with these entities, including the nature of accretion disks and shadows for instance. In essence, our attention is directed towards comprehensively investigating the behavior dictated by the geodesic equation. To achieve this goal, we write
\ie
\frac{\mathrm{d}^{2}x^{\mu}}{\mathrm{d}s^{2}} + \Gamma\indices{^\mu_\alpha_\beta}\frac{\mathrm{d}x^{\alpha}}{\mathrm{d}s}\frac{\mathrm{d}x^{\beta}}{\mathrm{d}s} = 0, \label{geodesicfull}
\fe
where $s$ is an arbitrary affine parameter. This investigation leads to the four coupled partial differential equations, which can be expressed as follows:
\ie
\begin{split}
&\frac{\mathrm{d}}{\mathrm{d}s}t' = -\frac{r' t' \left(-\frac{2 M r}{\left(a^2+r^2\right)^{3/2}}-\frac{3 \sqrt{M} r \sqrt{a^2+r^2} \sqrt{2 a^2 r+r^3}}{\left(\left(a^2+r^2\right)^{3/2}\right)^{3/2}}+\frac{\sqrt{M} \left(2 a^2+3 r^2\right)}{\sqrt{\left(a^2+r^2\right)^{3/2}} \sqrt{2 a^2 r+r^3}}\right)}{\frac{2 M}{\sqrt{a^2+r^2}}+\frac{2 \sqrt{M} \sqrt{2 a^2 r+r^3}}{\sqrt{\left(a^2+r^2\right)^{3/2}}}-1},
\end{split}
\fe
\ie
\begin{split}
\frac{\mathrm{d}}{\mathrm{d}s}r' &= -\frac{\sqrt{M} \left(r'\right)^2 \left(-2 a^6-a^4 r^2+2 \sqrt{M} r \sqrt{a^2+r^2} \sqrt{\left(a^2+r^2\right)^{3/2}} \sqrt{2 a^2 r+r^3}+a^2 r^4\right)}{ \left(2 M \sqrt{\left(a^2+r^2\right)^{3/2}}+2 \sqrt{M} \sqrt{a^2+r^2} \sqrt{2 a^2 r+r^3}-\sqrt{a^2+r^2} \sqrt{\left(a^2+r^2\right)^{3/2}}\right)}\\
&\times \frac{1}{2 \left(a^2+r^2\right)^{3/2} \sqrt{2 a^2 r+r^3}} \\
& -\frac{1}{2} \left(\frac{2 M r}{\left(a^2+r^2\right)^{3/2}}+\frac{3 \sqrt{M} r \sqrt{a^2+r^2} \sqrt{2 a^2 r+r^3}}{\left(\left(a^2+r^2\right)^{3/2}\right)^{3/2}}-\frac{\sqrt{M} \left(2 a^2+3 r^2\right)}{\sqrt{\left(a^2+r^2\right)^{3/2}} \sqrt{2 a^2 r+r^3}}\right)\\
& \times t'^2 \left(-\frac{2 M}{\sqrt{a^2+r^2}}-\frac{2 \sqrt{M} \sqrt{2 a^2 r+r^3}}{\sqrt{\left(a^2+r^2\right)^{3/2}}}+1\right)\\
& -r \left(\theta '\right)^2 \left(\frac{2 M}{\sqrt{a^2+r^2}}+\frac{2 \sqrt{M} \sqrt{2 a^2 r+r^3}}{\sqrt{\left(a^2+r^2\right)^{3/2}}}-1\right) \\
& + r \sin ^2(\theta ) \left(\varphi '\right)^2 \left(-\frac{2 M}{\sqrt{a^2+r^2}}-\frac{2 \sqrt{M} \sqrt{2 a^2 r+r^3}}{\sqrt{\left(a^2+r^2\right)^{3/2}}}+1\right),
\end{split}
\fe
\ie
\frac{\mathrm{d}}{\mathrm{d}s}\theta' = \sin (\theta ) \cos (\theta ) \varphi'^2-\frac{2 r \theta ' r'}{a^2+r^2},
\fe
\ie
\frac{\mathrm{d}}{\mathrm{d}s}\varphi' = 2 \varphi ' \left(-\frac{r r'}{a^2+r^2}-\theta ' \cot (\theta )\right).
\fe
As evident from the derived geodesic equations, we encounter four coupled partial differential equations. These ones serve as the foundation for determining the behavior of photon particles. Consequently, we perform ray tracing simulations to visualize the paths of light, as depicted in Fig. \ref{raytracing}.

\section{Critical orbits and shadows}

A comprehensive understanding of particle dynamics and the behavior of light rays in the vicinity of black hole structures necessitates a profound knowledge of critical orbits. These orbits hold a pivotal significance in elucidating the properties of spacetime influenced by dark matter effects in our specific context.

To achieve a more profound comprehension of the photon sphere's influence, often referred to as the critical orbit, in our black hole scenario, we will utilize the Lagrangian method to calculate null geodesics. This approach provides a clearer and more accessible understanding for readers compared to the utilization of the previously presented geodesic equation. Through this analysis, we aim to explore how the black hole mass impacts the photon sphere, shedding light on the gravitational effects inherent in the Simpson--Visser solution within Verlinde's emergent gravity framework. In this sense, we write:
\begin{equation}
\mathcal{L} = \frac{1}{2} g_{\mu\nu}\Dot{x}^{\mu}\Dot{x}^{\nu}.
\end{equation}
Upon considering a fixed angle of $\theta=\pi/2$, the aforementioned expression undergoes simplification, resulting in:
\ie
g_{00}^{-1} E^{2} + g_{11}^{-1} \Dot{r}^{2} + g_{33}L^{2} = 0, \label{compositions}
\fe
with $L$ is the angular momentum and $E$ being the energy. Next, Eq. (\ref{compositions}) reads,
\ie
\Dot{r}^{2} = E^{2} - \left(  1 - \frac{2M}{\sqrt{r^{2}+a^{2}}} - 2\sqrt{M}\sqrt{\frac{r(r^{2}+2a^{2})}{(r^{2}+a^{2})^{3/2}}} \right)\left(  \frac{L^{2}}{(r+a)^{2}} \right),
\fe
with $\overset{\nsim}{V} \equiv \left(  1 - \frac{2M}{\sqrt{r^{2}+a^{2}}} - 2\sqrt{M}\sqrt{\frac{r(r^{2}+2a^{2})}{(r^{2}+a^{2})^{3/2}}} \right)\left(  \frac{L^{2}}{(r+a)^{2}} \right)$ is the effective potential. In order to ascertain the critical radius, we must solve the equation $\partial \overset{\nsim}{V}/\partial r = 0$. Here, considering $a$ small, there exist two physical solutions for this equation (when the following range is considered for the values of $M$: $[0.1,0.2]$), i.e., two photon spheres. These ones are represented by $r_{c_{-}}$ (the inner photon sphere) and $r_{c_{+}}$ (the outer photon sphere). It is worth mentioning that, although we have considered small values for $a$, all outputs will numerically be displayed.

In order to gain a deeper understanding of the behavior of $r_{c_{-}}$ and $r_{c_{+}}$, we present Tables \ref{criticalradius} and \ref{criticalradius2}. Examining the left side of Table \ref{criticalradius}, we observe that the inner photon sphere expands with increasing $a$ (for $M=0.1$). On the right side of the same table, we note that there exists an augmented radius for the critical orbit $r_{c_{-}}$ when the mass $M$ increases (for $a=0.1$). Turning our attention to Table \ref{criticalradius2}, it illustrates the values of $r_{c_{+}}$ for different parameters $M$ and $a$. Here, the increase of $a$ and $M$ corresponds to a higher value of $r_{c_{+}}$.

Recently in the literature, Ref. \cite{aa2023analysis} addressed a similar study also in this context of Verlinde's emergent gravity. Additionally, it is important to mention that the appearance of two photon spheres instead was recently reported within the context Simpson--Visser solution \cite{tsukamoto2021gravitational,tsukamoto2022retrolensing} and others \cite{guerrero2022multiring}.

\begin{table}[!h]
\begin{center}
\begin{tabular}{c c  c ||| c c c } 
 \hline\hline
 $M$ & $a$ & $r_{c_{-}}$ & $M$ & $a$ & $r_{c_{-}}$  \\ [0.2ex] 
 \hline 
  0.1 & 0.10 & 0.240199 & 0.10 & 0.10 & 0.240199  \\ 

  0.1 & 0.11 & 0.259689 & 0.11 & 0.10 & 0.244980  \\
 
  0.1 & 0.12 & 0.277993 & 0.12 & 0.10 & 0.248546  \\
 
  0.1 & 0.13 & 0.295043 & 0.13 & 0.10 & 0.251283  \\
 
  0.1 & 0.14 & 0.310794 & 0.14 & 0.10 & 0.253437  \\
 
  0.1 & 0.15 & 0.325227 & 0.15 & 0.10 & 0.255169  \\
 
  0.1 & 0.16 & 0.338358 & 0.16 & 0.10 & 0.256588  \\
 
  0.1 & 0.17 & 0.350226 & 0.17 & 0.10 & 0.257768   \\
 
 0.1 & 0.18 & 0.360897 & 0.18 & 0.10 & 0.258762    \\
 
 0.1 & 0.19 & 0.370456 & 0.19 & 0.10 & 0.259611    \\
 
 0.1 & 0.20 & 0.378996 & 0.20 & 0.10 & 0.260342    \\
 [0.2ex] 
 \hline \hline
\end{tabular}
\caption{\label{criticalradius} The values of the critical orbits $r_{c_{-}}$ are displayed for different values of mass $M$ and parameter $a$. This table was entirely modified}
\end{center}
\end{table}

\begin{table}[!h]
\begin{center}
\begin{tabular}{c c  c ||| c c c } 
 \hline\hline
 $M$ & $a$ & $r_{c_{+}}$ & $M$ & $a$ & $r_{c_{+}}$  \\ [0.2ex] 
  \hline 
  0.1 & 0.10 & 0.602506 & 0.10 & 0.10 & 0.602506  \\ 

  0.1 & 0.11 & 0.613016 & 0.11 & 0.10 & 0.676538  \\
 
  0.1 & 0.12 & 0.624712 & 0.12 & 0.10 & 0.757195  \\
 
  0.1 & 0.13 & 0.637662 & 0.13 & 0.10 & 0.844476  \\
 
  0.1 & 0.14 & 0.651912 & 0.14 & 0.10 & 0.938567  \\
 
  0.1 & 0.15 & 0.667478 & 0.15 & 0.10 & 1.039800  \\
 
  0.1 & 0.16 & 0.684348 & 0.16 & 0.10 & 1.148600  \\
 
  0.1 & 0.17 & 0.702479 & 0.17 & 0.10 & 1.265530   \\
 
 0.1 & 0.18 & 0.721808 & 0.18 & 0.10 & 1.391240    \\
 
 0.1 & 0.19 & 0.742249 & 0.19 & 0.10 & 1.526470    \\
 
 0.1 & 0.20 & 0.763709 & 0.20 & 0.10 & 1.672110    \\[0.2ex] 
 \hline \hline
\end{tabular}
\caption{\label{criticalradius2} The values of the critical orbits $r_{c_{+}}$ are displayed for different values of mass $M$ and parameter $a$. }
\end{center}
\end{table}

Furthermore, the study of shadows in the context of emergent gravity and black hole structures holds immense significance as it offers a remarkable perspective of understanding the fundamental properties of these objects. Shadows, formed by the apparent silhouette of a black hole against the surrounding bright background, carry valuable features about the spacetime geometry and gravitational effects in the vicinity of the black hole. Analyzing and characterizing these shadows can provide fruitful knowledge into testing and refining theoretical models, verifying the nature of gravity. To do so, in order to facilitate our analysis, we conveniently introduce two new parameters, namely,
\begin{equation}\label{par}
\xi  = \frac{L}{E}
\text{ and }
\eta  = \frac{\mathcal{K}}{{E^2}},
\end{equation}
where $\mathcal{K}$ is the Carter constant. After some algebraic manipulations, we get
\begin{equation}
{\xi ^2} + \eta  =  \frac{r_{c_{\pm}}^2}{f(r_{c_{\pm}})}.
\end{equation}

In our quest to determine the radius of the shadow, we will employ the celestial coordinates $\alpha$ and $\beta$ \cite{singh2018shadow,heidari2023gravitational,hassanabadi2023gravitational} in the following manner: $\alpha=-\xi$ and $\beta=\pm\sqrt{\eta}$. Utilizing these coordinates, we obtain the shadow radius:

\begin{equation}
\begin{split}
\mathcal{R}_{\pm} = \frac{r_{c_{\pm}}}{{\sqrt{|f(r_{c_{\pm}})|}}} .
\end{split}
\end{equation}

\begin{figure}
    \centering
    \includegraphics[scale=0.34]{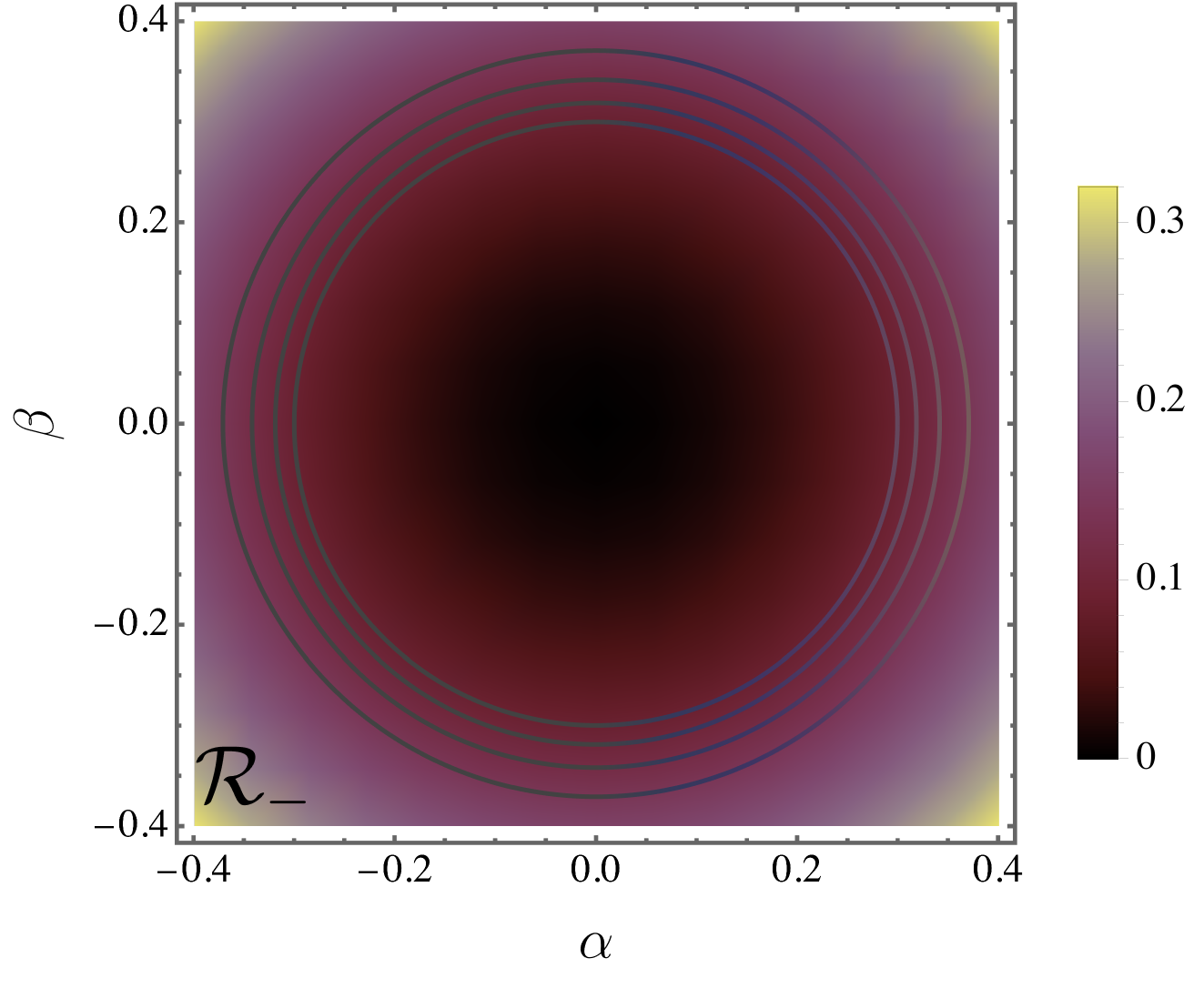}
    \includegraphics[scale=0.34]{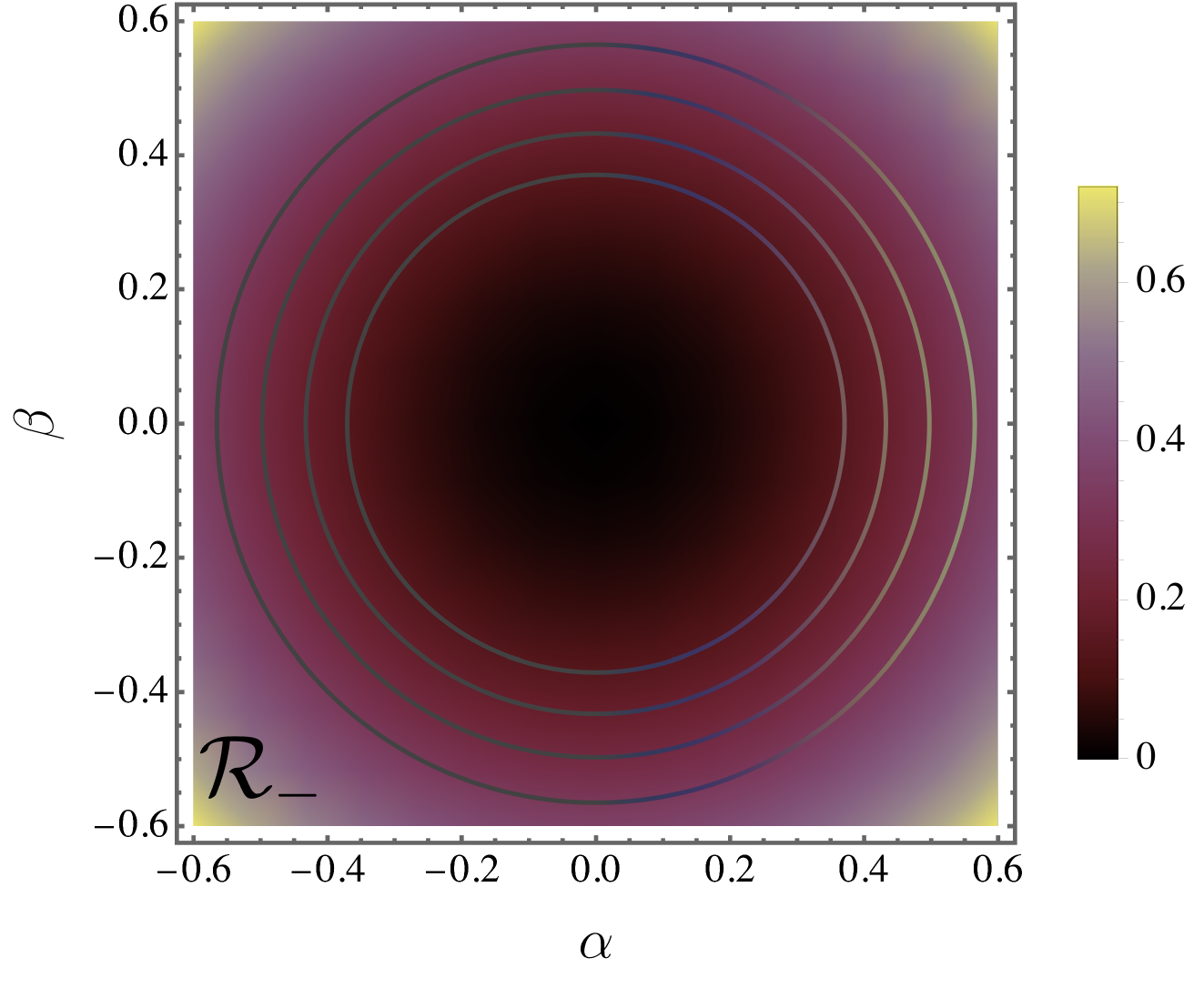}
    
    \caption{The circles in the illustration depict the shadows cast by varying values of mass $M$ and $a$, corresponding to the photon sphere $r_{c_{-}}$. Specifically, on the left side, the outer radius corresponds to $M=0.10$, followed by $M=0.11$, $M=0.12$, and the inner radius at $M=0.13$, all for $a=0.1$. Furthermore, on the right side, we explore values for $a$ at $M=0.10$, ranging from the innermost at $a=0.01$ to the outermost at $a=0.13$, i.e., the values regarded to this latter case are $a=0.01$, $a=0.11$, $a=0.12$, and $a=0.13$.}
    \label{shadows2}
\end{figure}

In Fig. \ref{shadows2}, the illustration features circles that represent the shadows cast by varying mass $M$ and $a$ associated with the photon sphere $r_{c_{-}}$. To elaborate, on the left side, the sequence begins with the outer radius at $M=0.10$, followed by $M=0.11$, $M=0.12$, and culminating with the inner radius at $M=0.13$, all under the condition $a=0.10$. Here, it can be observed that as the mass $M$ increases, the shadow radius decreases. This interesting phenomenon illustrates the gravitational effect of the dark matter on the surrounding space, causing a noticeable reduction in the apparent size of the shadow when $M$ increases. Moving to the right side, we examine different values of $a$ for $M=0.10$, ranging from the innermost at $a=0.10$ to the outermost at $a=0.13$. In other words, the considered values for this latter scenario encompass $a=0.10$, $a=0.11$, $a=0.12$, and $a=0.13$. Naturally, it is inherently verifiable that with the augmentation of the parameter $a$, there is a concomitant increase in the magnitude of the cast shadows.

\begin{figure}
    \centering
    \includegraphics[scale=0.34]{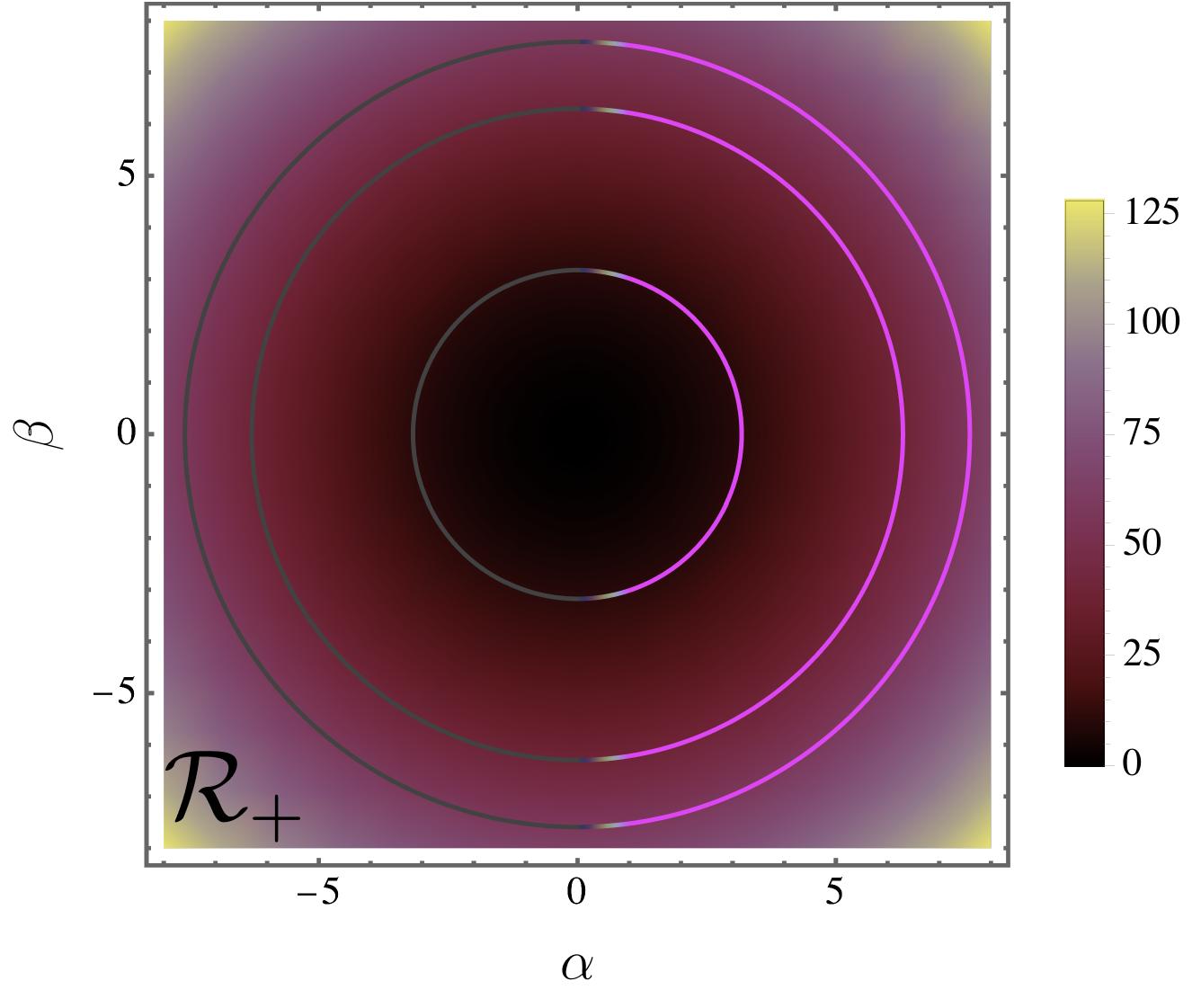}
    \includegraphics[scale=0.34]{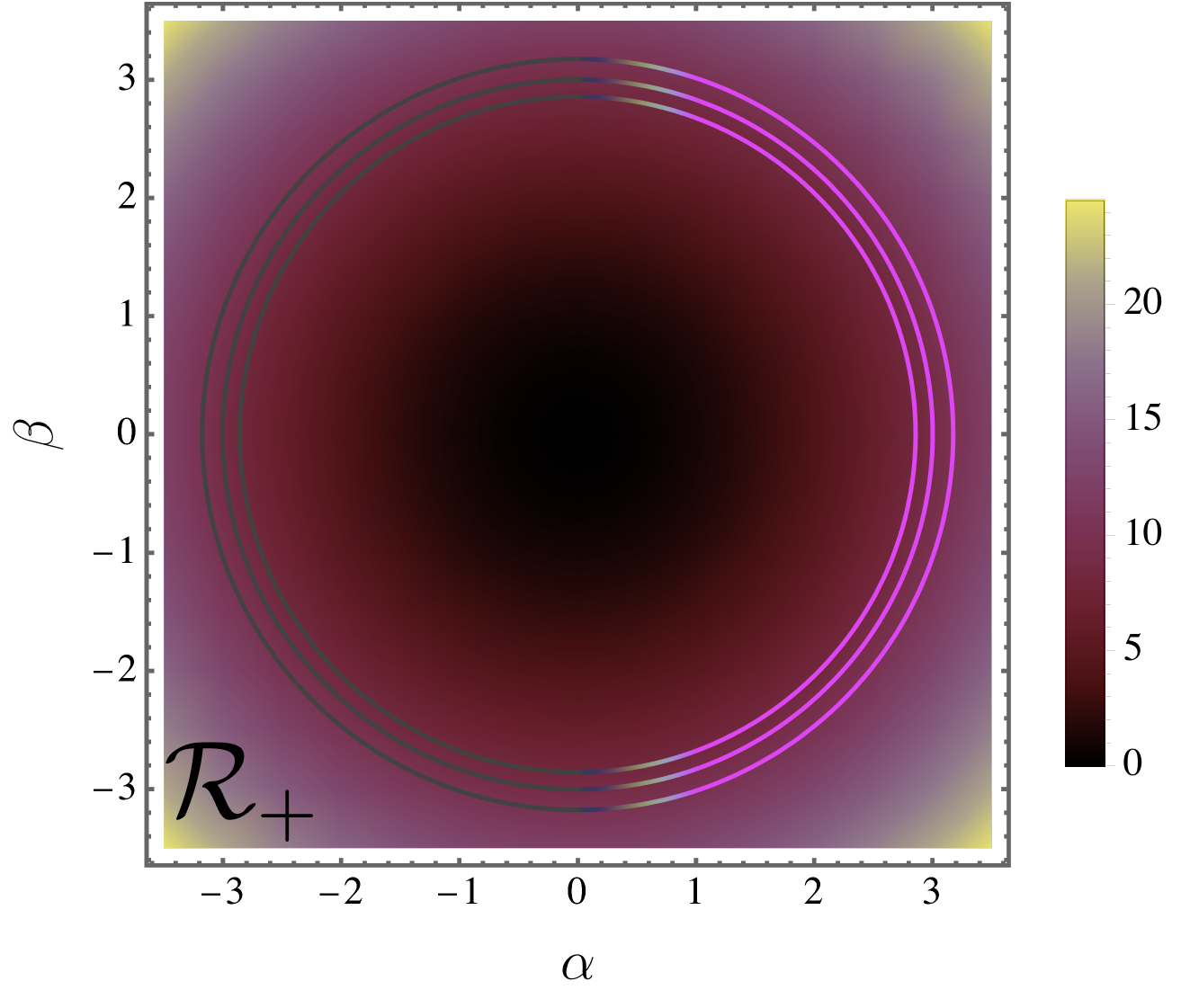}
    
    \caption{The circles symbolize shadows cast by varying masses $M$ corresponding to the photon sphere $r_{c_{+}}$. To elaborate, on the left side, the sequence progresses from the inner radius at $M=0.10$, followed by $M=0.11$, culminating with the outer radius at $M=0.12$, all with a fixed parameter $a=0.1$. Conversely, on the right side, we systematically explore different values of $a$ for a constant mass $M=0.10$, ranging from the outermost at $a=0.10$ to the innermost at $a=0.12$. More explicitly, the values considered for this latter scenario encompass $a=0.10$, $a = 0.11$, and $a=0.12$.}
    \label{shadows1}
\end{figure}

In Fig. \ref{shadows1}, the diagram depicts circles symbolizing shadows cast by varying mass $M$ and a associated with the photon sphere $r_{c_{+}}$. On the left side, the sequence initiates with the inner radius at $M=0.10$, followed by $M=0.11$, and concluding with the outer radius at $M=0.12$, all within the confines of the condition $a=0.1$. This noteworthy phenomenon underscores the gravitational impact of dark matter on the surrounding space, inducing a perceptible increment in the apparent size of the shadow as $M$ increases for a fixed value of $a$, i.e., $a=0.10$, and for the following range of $M$: $[0.10,0.12]$.

Transitioning to the right side, we scrutinize various values of $a$ for $M=1$, spanning from the outermost at $a=0.10$ to the innermost at $a=0.12$. Specifically, the values considered in this latter scenario encompass $a=0.10$, $a=0.11$, and $a=0.12$. Inevitably, it is demonstrably evident that an increase in the parameter $a$ is intricately associated with a simultaneous reduction in the magnitude of the cast shadows for $\mathcal{R}_{+}$. These results are in contrast to those ones encountered in the analysis of $\mathcal{R}_{-}$.

Furthermore, the quantitative values of $\mathcal{R}_{\pm}$ are displayed in Tabs. \ref{shaddowwsminus} and \ref{shaddowwsplus} for a better comprehension to the reader.

\begin{table}[!h]
\begin{center}
\begin{tabular}{c c  c ||| c c c } 
 \hline\hline
 $M$ & $a$ & $\mathcal{R}_{-}$ & $M$ & $a$ & $\mathcal{R}_{-}$  \\ [0.2ex] 
 \hline 
   
  0.10 & 0.10 & 0.370765 & 0.10 & 0.10 & 0.370765  \\
 
  0.10 & 0.11 & 0.432468 & 0.11 & 0.10 & 0.341788  \\
 
  0.10 & 0.12 & 0.497373 & 0.12 & 0.10 & 0.318724  \\
 
  0.10 & 0.13 & 0.565027 & 0.13 & 0.10 & 0.299818  \\
 
  0.10 & 0.14 & 0.634945 & 0.14 & 0.10 & 0.283968  \\
 
  0.10 & 0.15 & 0.706631 & 0.15 & 0.10 & 0.270440  \\
 
  0.10 & 0.16 & 0.779662 & 0.16 & 0.10 & 0.258724   \\
 
 0.10 & 0.17 & 0.853679 & 0.17 & 0.10 & 0.248450    \\
 
 0.10 & 0.18 & 0.928465 & 0.18 & 0.10 & 0.239345    \\
 
 0.10 & 0.19 & 1.003970 & 0.19 & 0.10 & 0.231207   \\
 
  0.10 & 0.20 & 1.080290 & 0.20 & 0.10 & 0.223874   \\
 [0.2ex] 
 \hline \hline
\end{tabular}
\caption{\label{shaddowwsminus} The values of shadows $\mathcal{R}_{-}$ are displayed for different values of mass $M$ and parameter $a$. }
\end{center}
\end{table}

\begin{table}[!h]
\begin{center}
\begin{tabular}{c c  c ||| c c c } 
 \hline\hline
 $M$ & $a$ & $\mathcal{R}_{+}$ & $M$ & $a$ & $\mathcal{R}_{+}$  \\ [0.2ex] 
 \hline 
  
  0.10 & 0.10 & 3.17571 & 0.10 & 0.10 & 3.17571  \\
 
  0.10 & 0.11 & 3.00205 & 0.11 & 0.10 & 6.29674  \\
 
  0.10 & 0.12 & 2.85755 & 0.12 & 0.10 & 7.59084  \\
 
  0.10 & 0.13 & 2.73754 & 0.13 & 0.10 & 4.93324  \\
 
  0.10 & 0.14 & 2.63852 & 0.14 & 0.10 & 4.32712  \\
 
  0.10 & 0.15 & 2.55774 & 0.15 & 0.10 & 4.12499  \\
 
  0.10 & 0.16 & 2.49291 & 0.16 & 0.10 & 4.08552   \\
 
 0.10 & 0.17 & 2.44209 & 0.17 & 0.10 & 4.13394    \\
 
 0.10 & 0.18 & 2.40355 & 0.18 & 0.10 & 4.23917    \\
 
 0.10 & 0.19 & 2.37575 & 0.19 & 0.10 & 4.38623   \\
 
  0.10 & 0.20 & 2.35733 & 0.20 & 0.10 & 4.56765   \\
 [0.2ex] 
 \hline \hline
\end{tabular}
\caption{\label{shaddowwsplus} The values of shadows $\mathcal{R}_{+}$ are displayed for different values of mass $M$ and parameter $a$. }
\end{center}
\end{table}

\section{Time delay and Deflection angle}

The examination of time delay in dark matter scenarios holds profound significance within theoretical physics. This exploration provides profound insights into the fundamental aspects of a Simpson-Visser solution in Verlinde gravity. To determine the time delay, we employ the Lagrangian for null geodesics, enabling an accurate quantification of the temporal delay experienced by particles. This sheds light on the intricate dynamics of such a spacetime, as expressed by the equation:

\begin{equation}
	\left(\frac{{\mathrm{d}r}}{{\mathrm{d}t}}\right)^2 = f(r)^2  \left(1-\frac{ r_{\min }^2 f(r)}{r^2 f\left(r_{\min }\right)}\right)
\end{equation}

Also, knowing that $\mathrm{d}r/\mathrm{d}t=0$ at $r=r_{\min}$, we derive:

\begin{equation}
	\frac{{\mathrm{d}t}}{{\mathrm{d}r}} = \frac{1}{{{f(r) }\sqrt {1 - {{(\frac{{{r_{\min }}}}{r})}^2}\frac{{{f }(r)}}{{{f }({r_{\min }})}}} }}.
\end{equation}

Thereby, the time delay can be computed as 
\ie
\begin{split}
t(r,r_{\text{min}})  = & \int^{\infty}_{r_{\text{min}}} \frac{\mathrm{d}r}{{{f(r) }\sqrt {1 - {{(\frac{{{r_{\min }}}}{r})}^2}\frac{{{f }(r)}}{{{f }({r_{\min }})}}} }} \\
= & \int^{\infty}_{r_{\text{min}}}  \frac{\mathrm{d}r}{\left(-\frac{2 M}{\sqrt{a^2+r^2}}-\frac{2 \sqrt{M} \sqrt{2 a^2 r+r^3}}{\sqrt{\left(a^2+r^2\right)^{3/2}}}+1\right) \sqrt{1-\frac{r_{\min }^2 \left(-\frac{2 M}{\sqrt{a^2+r^2}}-\frac{2 \sqrt{M} \sqrt{2 a^2 r+r^3}}{\sqrt{\left(a^2+r^2\right)^{3/2}}}+1\right)}{r^2 \left(-\frac{2 M}{\sqrt{a^2+r_{\min }^2}}-\frac{2 \sqrt{M} \sqrt{2 a^2 r_{\min }+r_{\min }^3}}{\sqrt{\left(a^2+r_{\min }^2\right){}^{3/2}}}+1\right)}}}.
\end{split}
\fe

Furthermore, the bending of light as it traverses the curves of spacetime stands as a fundamental and intriguing phenomenon, serving as a crucial tool in the field of scientific inquiry. Light, exhibiting an elegant conformity to the contours of spacetime, reveals a captivating interplay that imparts invaluable understanding into the intricate physics governing gravitational sources \cite{170,171,172,173,174,175}.

To quantitatively describe this deflection angle, we employ a well--established formula \cite{76,jusufi2020quasinormal} expressed as:

\begin{equation}\label{phi}
	\hat \alpha ({r_{\min }}) = 2\int_{{r_{\min }}}^\infty  {\frac{\mathrm{d}r}{{r\sqrt {{{(\frac{r}{{{r_{\min }}}})}^2}{f }({r_{\min }}) - {f }(r)} }} - \pi } .
\end{equation}

In our analysis, we use the symbol $\hat{\alpha}$ to denote the deflection angle.  Our approach entails the initial determination of the minimum radius, $r_{\text{min}}$, along the trajectory corresponding to each specific $a$ and $M$. Particularly, in Tab. \ref{Table3ed}, we display the deflection angle $\Hat{\alpha}$, considering different values of $a$ with a fixed $M$. Here, we notice a very small difference between them. Such an aspect is expected, since we have been dealing with small $a$.

\begin{table}[!ht]
	\centering
	\caption{Variation of $r_{min}$ and the deflection angle for various $a$ and impact parameter ($L/E$) $=10$ (when $M=0.1$).}
	\begin{tabular}{|c|c|c|c|c|c|c|c|c|c|c|c|}
\hline
$M$ & 0.10 & 0.10 & 0.10 & 0.10 & 0.10 & 0.10 & 0.10 & 0.10 & 0.10 & 0.10 & 0.10 \\		
  \hline
		$a$ & 0.00 & 0.01 & 0.02 & 0.03 & 0.04 & 0.05 & 0.06 & 0.07 & 0.08 & 0.09 & 0.10 \\ \hline
		$r_{min}$ & 0.544152 & 0.544139 & 0.544100 & 0.544033 & 0.543936 & 0.543805 & 0.543636 & 0.543423 & 0.54316 & 0.542841 & 0.542458  \\ \hline
		$\hat{\alpha}$ & 2.38338 & 2.38339 & 2.38341 & 2.38343 & 2.38347 & 2.38352 & 2.38359 & 2.38366 & 2.38374 & 2.38384 & 2.38395 \\ \hline
	\end{tabular}
	\label{Table3ed}
\end{table}


\section{Neutrino Energy Deposition in High--Energy Astrophysical Phenomena}

The study is focused on examining the energy deposition rate resulting from the $\nu\nu^{-} \rightarrow e+e^{-}$ process, with the aim of elucidating its relevance to \textit{gamma ray burst} emissions. The investigated scenario pertains to the concluding phase of neutron star mergers, conceptualized as a black hole accompanied by an accretion disk.

Salmonson and Wilson, as highlighted in Ref. \cite{73,74}, pioneered the exploration of the impacts within strong gravitational field regimes. Their groundbreaking work revealed that, within a Schwarzschild spacetime and for neutrinos emitted from the central core, the efficiency of the annihilation process $\nu\nu^{-} \rightarrow e+e^{-}$ undergoes substantial amplification—approximately 30 times greater than its Newtonian counterpart--especially in the context of collapsing neutron stars.

Building upon this foundation, subsequent studies \cite{75,076} further examined the ramifications of general relativity on neutrino pair annihilation. This exploration extended to regions near the neutrinosphere and in proximity to a thin accretion disk (assuming an isothermal profile), with the gravitational background characterized by both Schwarzschild and Kerr geometries.

Our analysis centers around a black hole (BH) encircled by a slender accretion disk that emits neutrinos, as elaborated in \cite{076}. We concentrate on an idealized model, free from reliance on the specifics of disk formation and excluding considerations of self-gravitational effects. This disk is characterized by well--defined inner and outer edges, represented by radii denoted as $R_{in}$ and $R_{out}$, respectively \cite{lambiase2023probing}.

The Hamiltonian proves instrumental in analyzing the trajectory of a test particle within spacetime. It facilitates the computation of crucial parameters, including the energy and angular momentum of the test particle, and allows for the derivation of its equations of motion. In the scenario of a test particle traversing a curved background, the Hamiltonian takes the form
\ie 
2 \mathcal{H} = -E\dot{t} + L\dot{\phi} + g_{rr}\dot{r}^2 = 0.
\fe
Also, the non-zero components of the $4$--velocity can be obtained as follows:
\ie
U^3 = \dot{\phi} = -\frac{L}{r^2}, \nonumber
\fe
\ie
U^0 = \dot{t} = -\frac{E}{g_{tt}}, \nonumber
\fe
\ie
\Dot{r}^{2} = \frac{E \Dot{t} - L \Dot{\phi}}{g_{11}}  .
\fe

Our emphasis is on determining the rate of energy deposition in close proximity to the axis, which is perpendicular to the disk, particularly at $\theta = 0^{\text{o}}$. To evaluate the energy emitted within a half cone with an angular extent of approximately $\Delta \theta \approx 10^{\text{o}}$, we must consider the scalar product of the momenta of a neutrino and an antineutrino at $\theta = 0^{\text{o}}$. This scalar product can be expressed as 
\ie
p_\nu \cdot p_{\bar{\nu}} = E_\nu E_{\bar{\nu}} \left\{1 - \sin\theta_\nu \sin\theta_{\bar{\nu}} \cos(\phi_\nu - \phi_{\bar{\nu}}) - \cos\theta_\nu \cos\theta_{\bar{\nu}} \right\}.
\fe

Here, the term \(E_\nu\) is defined as the energy of the neutrino, calculated as \(E_0^\nu/\sqrt{g_{00}}\). Also, \(E_0^\nu\) is defined as the observer energy of the neutrino measured at infinity and
\ie
\sin\theta_\nu = \frac{\rho_\nu}{\sqrt{-1 + \frac{2M}{\sqrt{r^{2}+a^{2}}} + 2\sqrt{M}\sqrt{\frac{r(r^{2}+2a^{2})}{(r^{2}+a^{2})^{3/2}}}}}.
\fe
In this context, \(\rho_{\nu}\) is precisely defined as the ratio of the angular momentum \(L_{\nu}\) to the observed energy \(E_{0\nu}\). Moreover, considering geometric factors, there exist both a minimum and maximum value, denoted as $\theta_{m}$ and $\theta_{M}$ respectively, for a neutrino originating from $R_{in} = 2 r_{c}$ to $R_{out} = 30M$, where $r_{c}$ denotes the radius of the photosphere. Furthermore, it can be shown that the following relationship holds \cite{076}:
\ie
\rho_{\nu} = \frac{r_{0}}{\sqrt{-1 + \frac{2M}{\sqrt{r_{0}^{2}+a^{2}}} + 2\sqrt{M}\sqrt{\frac{r_{0}(r_{0}^{2}+2a^{2})}{(r_{0}^{2}+a^{2})^{3/2}}}}},
\fe
where, \(r_0\) represents the closest distance between the particle and the center before reaching $\theta = 0$. The ultimate component is the trajectory equation, written as \cite{lambiase2023probing}
\ie
\int \frac{\mathrm{d}r^{\prime}}{r^{\prime} \sqrt{\left(\frac{r^{\prime}}{\rho_{\nu}}\right)^{2} +1 - \frac{2M}{\sqrt{r^{2}+a^{2}}} - 2\sqrt{M}\sqrt{\frac{r(r^{2}+2a^{2})}{(r^{2}+a^{2})^{3/2}}}}} = \frac{\pi}{2}.
\fe

The equation above considers the emission of neutrinos from the position \((R, \pi/2)\), where \(R\) ranges between \(R_{\text{in}}\) and \(R_{\text{out}}\). These neutrinos then travel to the position \((r, 0)\). As a result, the energy deposition rate resulting from neutrino pair annihilation is elaborated upon in \cite{076}
\ie
\frac{\mathrm{d}E_{0}(r)}{\mathrm{d}t \mathrm{d}V}  = \frac{21 \pi^4}{4}\zeta(5)K G_{\mathfrak{F}}^2 k^9T^9_{\text{eff}} (2 r_{c})F(r),
\fe
where \(k\) stands for the Boltzmann constant, $G_{\mathfrak{F}}$ represents the Fermi constant,  \(T_{\text{eff}}(2r_{c})\) denotes the effective temperature at a radius of $2r_{c}$, and $\zeta(s)$ is the Riemann zeta function as follows
\ie
\zeta(s) = \sum^{\infty}_{n=1} \frac{1}{n^{s}} = \frac{1}{\Gamma(s)}\int_{0}^{\infty} \frac{x^{s-1}}{e^{x}-1} \mathrm{d}x, \,\,\,\,\,\,\, \Gamma(s)= \int_{0}^{\infty} x^{s-1} e^{-x} \mathrm{d}x.
\fe

After an algebraic manipulation, we get
\ie
\begin{split}
F(r) =&  \frac{2 \pi^{2}}{T^{9}_{\text{eff}}(2 r_{c})} \frac{1}{\left( -1 + \frac{2M}{\sqrt{r^{2}+a^{2}}} + 2\sqrt{M}\sqrt{\frac{r(r^{2}+2a^{2})}{(r^{2}+a^{2})^{3/2}}} \right)^{4}} \\
\times & \left(   2 \int^{\theta M}_{\theta m} \mathrm{d} \theta_{\nu} T^{5}_{0} (\theta_{\nu}) \sin\theta_{\nu} \int^{\theta M}_{\theta m}  \mathrm{d}\theta_{\Bar{\nu}} T^{4}_{0}(\theta_{\Bar{\nu}})\sin\theta_{\Bar{\nu}} \right. \\
& \left. + \int^{\theta_{M}}_{\theta_{m}} \mathrm{d}\theta_{\nu} T^{5}_{0}(\theta_{\nu}) \sin^{3}\theta_{\nu} \int^{\theta_{M}}_{\theta_{m}} \mathrm{d}\theta_{\Bar{\nu}}T^{4}_{0}(\theta_{\Bar{\nu}})\sin^{3}\theta_{\Bar{\nu}}   \right. \\
& \left.   + 2 \int^{\theta_{M}}_{\theta_{m}} \mathrm{d} \theta_{\nu} T^{5}_{0}(\theta_{\nu}) \cos^{2}\theta_{\nu}\sin\theta_{\nu}  \int^{\theta_{M}}_{\theta_{m}} \mathrm{d} \theta_{\Bar{\nu}} T^{4}_{0}(\theta_{\Bar{\nu}}) \cos^{2}\theta_{\Bar{\nu}}\sin\theta_{\Bar{\nu}}   \right. \\
& \left.    -4 \int^{\theta_{M}}_{\theta_{m}} \mathrm{d} \theta_{\nu} T^{5}_{0}(\theta_{\nu}) \cos\theta_{\nu}\sin\theta_{\nu}  \int^{\theta_{M}}_{\theta_{m}} \mathrm{d} \theta_{\Bar{\nu}} T^{4}_{0}(\theta_{\Bar{\nu}}) \cos\theta_{\Bar{\nu}}\sin\theta_{\Bar{\nu}}                        \right),
\end{split}
\fe
where $T_{\text{eff}}$ denotes the effective temperature as measured by a local observer and $T_{0}$ represents the temperature observed at infinity, having its general form is given by
\ie
T_{0} = \frac{T_{\text{eff}}(R)}{\gamma} \sqrt{g_{00}(R)} = \frac{T_{\text{eff}}(R)}{\gamma} \sqrt{-1 + \frac{2M}{\sqrt{R^{2}+a^{2}}} + 2\sqrt{M}\sqrt{\frac{R(R^{2}+2a^{2})}{(R^{2}+a^{2})^{3/2}}}},
\fe
with $\gamma = \frac{1}{\sqrt{1-\frac{v^{2}}{c^{2}}}}$ and
\ie
\frac{v^{2}}{c^{2}} = \frac{r \sin ^2(\theta ) \left(\frac{\sqrt{M} \sqrt{\left(a^2+r^2\right)^{3/2}} \left(2 a^4-a^2 r^2\right)}{\left(a^2+r^2\right)^{5/2} \sqrt{2 a^2 r+r^3}}-\frac{2 M r}{\left(a^2+r^2\right)^{3/2}}\right)}{2 \left(\frac{2 M}{\sqrt{a^2+r^2}}+\frac{2 \sqrt{M} \sqrt{2 a^2 r+r^3}}{\sqrt{\left(a^2+r^2\right)^{3/2}}}-1\right)}.
\fe

It is important to mention that all quantities are assessed at $\theta = \pi/2$. In this analysis, we do not account for the reabsorption of deposited energy by the black hole. Consequently, our focus centers on a scenario featuring a straightforward temperature gradient \cite{076}
\ie
T_{\text{eff}}(R) \sim \frac{2 r_{c}}{r}.
\fe
Moreover, the assumptions pertaining to temperature values and the configuration of the gradient model align with recent discoveries in neutrino--cooled accretion disk models, as shown in \cite{78,79,80}.

Anticipated within the theoretical framework is an effective maximum temperature, denoted as $T_{\text{eff}}$, typically falling in the order of $\mathcal{O}(10$ MeV). This magnitude assumes pivotal significance in achieving the observed neutrino disk luminosity, thereby rendering the disk luminosity comparatively invariant across diverse models. Given our non--engagement in numerical simulations, we adopt the assumption $T_{\text{eff}} \sim \mathcal{O}$($10$ MeV) to ensure a standardized basis for comparing the impacts of distinct gravitational models under identical conditions.

It is imperative to underscore that despite these theoretical suppositions, the exact temperature profile can only be ascertained through a comprehensive disk simulation originating from the merging of neutron stars with a meticulously defined geometry. Here, we set $G(r) = \frac{F(r)r^2}{4M^2}
$. Furthermore, $G(r)$ plays a pivotal role in computing the energy deposition rate and, consequently, in determining the energy deposition for a \textit{gamma ray burst} event. We calculate such an energy deposition rate within an infinitesimal angle $\mathrm{d}\theta$, considering a characteristic angle of $10^{\circ}$ degrees and a temperature of $10$ MeV \cite{lambiase2023probing,076}
\ie
\frac{\mathrm{d}E_{0}}{\mathrm{d}t} \simeq     4.41 \times 10^{48} \left( \frac{\Delta \theta }{10^{\circ}} \right)^{2} \left( \frac{k T_{\text{eff}}(R_{in})}{10 \text{MeV}} \right)^{9} \left( \frac{2 M}{10 \text{km}} \right) \int^{R_{\text{out}}}_{R_{\text{in}}} \mathrm{d}r\frac{G(r)}{2M} \,\,\,\text{erg s}^{-1}.
\fe


\section{The quasinormal modes}

Throughout the ringdown phase, a remarkable phenomenon known as \textit{quasinormal} modes emerges, displaying distinct oscillation patterns that remain not affected by the initial perturbations. Indeed, these modes manifest the intrinsic characteristics of the system and originate from the innate oscillations of spacetime, independent of specific initial conditions.

Unlike \textit{normal} modes, which pertain to closed systems, \textit{quasinormal} modes are associated with open systems. As a result, these modes dissipate energy gradually through the emission of gravitational waves. Mathematically, they can be characterized as poles of the complex Green function.

For determining their frequencies, one needs to find solutions to the wave equation within a system governed by a background metric $g_{\mu\nu}$. However, acquiring analytical solutions for these modes is often a challenging procedure.

\begin{figure}
    \centering
    \includegraphics[scale=0.4]{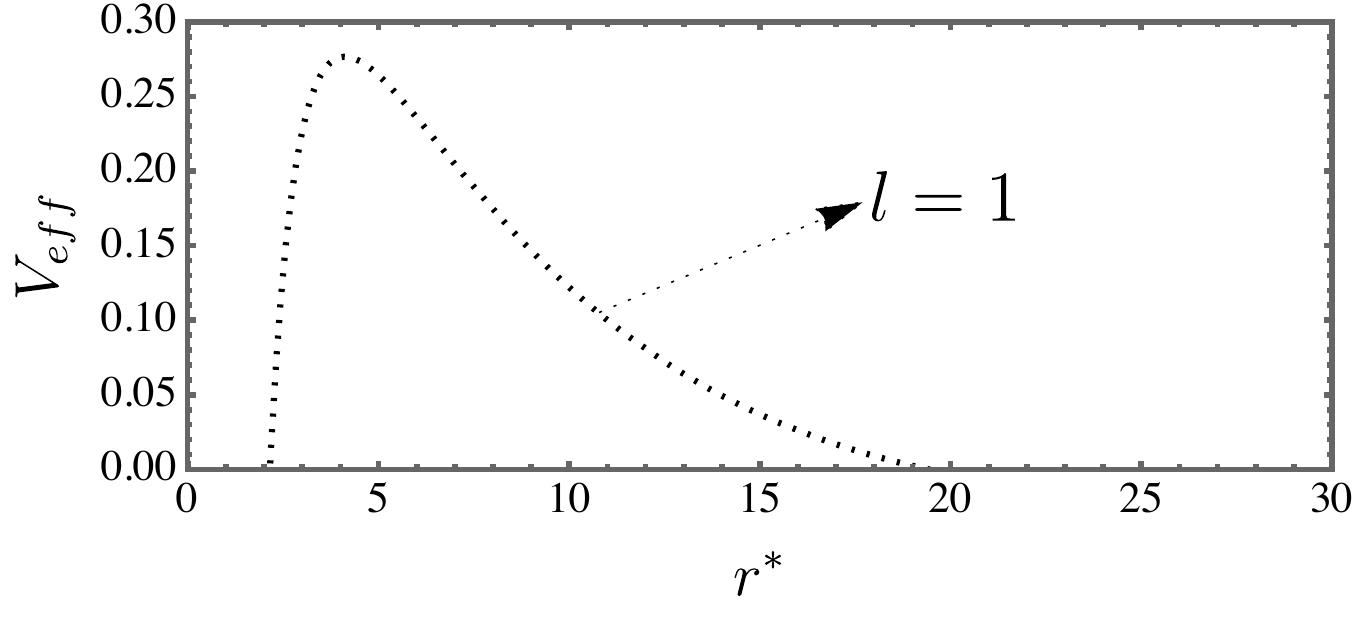}
    \caption{The effective potential $V_{eff}$ is depicted as a function of the tortoise coordinate $r^{*}$, specifically considering a particular value of $l$.}
    \label{effectivepotential}
\end{figure}

In the scientific literature, various techniques have been addressed to obtain solutions for these modes. Among them, the WKB (Wentzel--Kramers--Brillouin) approach stands out as one of the most widely used methods. Its development can be traced back to the groundbreaking work of Will and Iyer \cite{iyer1987black,iyer1987black1}, and subsequent advancements up to the sixth order were made by Konoplya \cite{konoplya2003quasinormal}. For our specific calculations, we focus on analyzing perturbations using the scalar field, which involves considering the Klein--Gordon equation within the context of a curved spacetime
\ie
\frac{1}{\sqrt{-g}}\partial_{\mu}(g^{\mu\nu}\sqrt{-g}\partial_{\nu}\Phi) = 0.\label{KL}
\fe
While the exploration of \textit{backreaction} effects in this particular scenario is intriguing, this manuscript does not provide this aspect and instead places its emphasis on other aspects. Specifically, our primary focus revolves around studying the scalar field as a small perturbation. Furthermore, the presence of spherical symmetry allows us to take advantage of this opportunity to decompose the scalar field in a specific manner, as elaborated below:
\ie
\Phi(t,r,\theta,\varphi) = \sum^{\infty}_{l=0}\sum^{l}_{m=-l}r^{-1}\Psi_{lm}(t,r)Y_{lm}(\theta,\varphi),\label{decomposition}
\fe
where we express the spherical harmonics as $Y_{lm}(\theta,\varphi)$. Also, we can substitute the decomposition of the scalar field, as shown in Eq. (\ref{decomposition}), into Eq. (\ref{KL}). This substitution transforms the equation into a Schrödinger--like form, endowed with wave--like properties, making it highly suitable for our analysis
\ie
-\frac{\partial^{2} \Psi}{\partial t^{2}}+\frac{\partial^{2} \Psi}{\partial r^{*2}} + V_{eff}(r^{*})\Psi = 0.\label{schordingereq}
\fe
The potential $V_{eff}$ is widely recognized as the \textit{Regge--Wheeler} potential or the effective potential, carrying crucial information about the black hole's geometry. Additionally, we introduce the tortoise coordinate $r^{*}$, which spans the entire spacetime as $r^{*}\rightarrow \pm \infty$. It is defined as $\mathrm{d} r^{*} = \sqrt{[1/f(r)^{2}]}\mathrm{d}r$.
After some algebraic manipulations, the effective potential reads:
\ie
\begin{split}
V_{eff}(r) = f(r) \left[\frac{\frac{3 \sqrt{M} r \sqrt{a^2+r^2} \sqrt{r \left(2 a^2+r^2\right)}}{\left(\left(a^2+r^2\right)^{3/2}\right)^{3/2}}+\frac{2 M r}{\left(a^2+r^2\right)^{3/2}}-\frac{\sqrt{M} \left(2 a^2+3 r^2\right)}{\sqrt{\left(a^2+r^2\right)^{3/2}} \sqrt{r \left(2 a^2+r^2\right)}}}{r}+\frac{l (l+1)}{r^2}\right]
\end{split}.
\fe
Figure \ref{effectivepotential} illustrates the effective potential $V_{eff}$ as a function of the tortoise coordinate $r^{*}$ for a particular value of $l$.

\subsection{The WKB approximation}

Here, our primary objective is to derive stationary solutions for the system, achieved by assuming that $\Psi(t,r)$ can be expressed as $\Psi(t,r) = e^{-i\omega t} \psi(r)$, where $\omega$ denotes the frequency. This assumption allows us to conveniently separate the time--independent component of Eq. (\ref{schordingereq}) by employing the following approach:
\ie
\frac{\partial^{2} \psi}{\partial r^{*2}} - \left[  \omega^{2} - V_{eff}(r^{*})\right]\psi = 0.\label{timeindependent}
\fe
To successfully solve Eq. (\ref{timeindependent}), careful consideration of the appropriate boundary conditions becomes crucial. In our specific case, the solutions that satisfy the necessary conditions are characterized by exhibiting purely ingoing behavior near the horizon
\[
    \psi^{\text{in}}(r^{*}) \sim 
\begin{cases}
    C_{l}(\omega) e^{-i\omega r^{*}} & ( r^{*}\rightarrow - \infty)\\
    A^{(-)}_{l}(\omega) e^{-i\omega r^{*}} + A^{(+)}_{l}(\omega) e^{+i\omega r^{*}} & (r^{*}\rightarrow + \infty).\label{boundaryconditions11}
\end{cases}
\]
In our analysis, the complex constants $C_{l}(\omega)$, $A^{(-)}_{l}(\omega)$, and $A^{(+)}_{l}(\omega)$ hold paramount significance. These ones are fundamental to explore the \textit{quasinormal} modes of a black hole, represented by frequencies $\omega_{nl}$ that satisfy the condition $A^{(-)}_{l}(\omega_{nl})=0$. Such modes exhibit a unique behavior, with purely outgoing waves at spatial infinity and purely ingoing waves at the event horizon. The integers $n$ and $l$ represent the overtone and multipole numbers. Also, it is worthy to be mentioned that the spectrum of \textit{quasinormal} modes is determined by the eigenvalues of Eq. (\ref{timeindependent}). To analyze these frequencies, we adopt the WKB method, i.e., a semi--analytical technique that draws parallels with quantum mechanics. This approach empowers us to gain valuable insights into the behavior of \textit{quasinormal} modes near the event horizon and at infinity, unraveling essential aspects of black hole dynamics and gravitational wave phenomena.

The WKB approximation, first introduced by Schutz and Will \cite{schutz1985black}, has become a valuable tool for computing \textit{quasinormal} modes in the context of particle scattering around black holes. Over time, this method has seen further refinements, notably by Konoplya \cite{konoplya2003quasinormal, konoplya2004quasinormal}. However, it is essential to note that the validity of this approach hinges on the potential exhibiting a barrier--like shape, approaching constant values as $r^{*} \rightarrow \pm \infty$. By fitting the power series of the solution near the turning points of the maximum potential, the \textit{quasinormal} modes can be reliably obtained \cite{santos2016quasinormal}. Then, the Konoplya formula reads:
\ie
\frac{i(\omega^{2}_{n}-V_{0})}{\sqrt{-2 V^{''}_{0}}} - \sum^{6}_{j=2} \Lambda_{j} = n + \frac{1}{2}.
\fe

Konoplya's formula for the \textit{quasinormal} modes, as mentioned earlier, incorporates several elements. The term $V^{''}_{0}$ represents the second derivative of the potential evaluated at its maximum point $r_{0}$, and $\Lambda_{j}$ are the constants that depend on the effective potential as well as of its derivatives at the maximum. It is worth noting that recent advancements in the field have introduced a 13th--order WKB approximation, proposed by Matyjasek and Opala \cite{matyjasek2017quasinormal}, which significantly improve the accuracy for the calculation of the \textit{quasinormal} frequencies.

Presented in Tables \ref{table5}, \ref{table6}, and \ref{table7}, we find a comprehensive compilation of \textit{quasinormal} frequencies obtained using the third--order WKB method. These tables are organized based on the multipole number $l$ and mass $M$. Notably, $\omega_{0}$, $\omega_{1}$, and $\omega_{2}$ for $M=0.1$, turn out to be unstable. Such behaviors may be attributed to the influence of dark matter, which is introduced in the initial conditions to yield a Simpson--Visser--like black hole solution.

It is crucial to emphasize that the \textit{quasinormal} modes linked to the scalar field exhibit a negative imaginary part. This significant characteristic implies that these modes experience exponential decay over time, signifying the dissipation of energy through scalar waves. This finding aligns with earlier investigations examining scalar, electromagnetic, and gravitational perturbations in spherically symmetric geometries \cite{ konoplya2011quasinormal,berti2009quasinormal,heidari2023gravitational,chen2023quasinormal}.

In a broader context, it becomes evident that as $M$ varies, there is an increase in the real part, coupled with a decrease in the imaginary part of the \textit{quasinormal} modes. This observation highlights the crucial role played by the mass parameter in governing the damping behavior of the scalar waves. Additionally, for all values of $l$ under consideration, the \textit{quasinormal} modes demonstrate a trend of exhibiting damper frequencies as $\omega_{n}$ increases.
It is important to mention that similar studies have been recently proposed in the literature, considering Hayward--like regular black holes \cite{konoplya2019higher,aa2023analysis}. Furthermore, Ref. \cite{akil2022semiclassical} analyzed a regular black--bounce model similar to the Simpson--Visser solution, deriving the thermodynamics, phase transition, \textit{Hawking} 
radiation, light ring, and the \textit{quasinormal} modes.

\begin{table}[!h]
\begin{center}
\begin{tabular}{c c c c} 
 \hline\hline
 $M$ & $\omega_{0}$ & $\omega_{1}$ & $\omega_{2}$  \\ [0.2ex] 
 \hline 
 0.1 & 0.106108 - 0.18710$i$ & 0.205563 - 0.519591$i$ & 0.376212 - 0.891811$i$ \\ 
 
 0.2 & 0.005412 - 0.00616$i$ & 0.00458125 - 0.0190045$i$ & 0.00318925 - 0.0318317$i$ \\
 
 0.3 & 4.96600$\times 10^{-14}$  - 4.96604$\times 10^{-14}$$i$ & 8.60138$\times10^{-14}$ - 8.60141$\times10^{-14}i$  &  1.11043$\times10^{-13}$ - 1.11044$\times 10^{-13}i$   \\
 
 0.4 & 1.23429$\times 10^{-13}$ - 1.23420$\times 10^{-13}i$ & 2.13786$\times10^{-13}$ - 2.13786$\times10^{-13}i$  &  2.75997$\times10^{-13}$ - 2.75997$\times10^{-13}i$  \\
 
 0.5 & 1.85834$\times10^{-8}$ -$ 1.85835\times10^{-8}i$ & 6.46340$\times10^{-8}$ - 6.46342$\times10^{-8}i$ & 1.32039$\times10^{-7}$ - 1.32039$\times10^{-7}i$  \\
 
 0.6 & 2.62198$\times10^{-13}$ - 2.62200$\times10^{-13}$$i$ & 4.54141$\times10^{-13}$ - 4.54142 $\times10^{-13} i$ & 5.86294$\times10^{-13}$ - 5.86294$\times10^{-13}i$ \\
 
 0.7 & 3.26881$\times10^{-8}$ - 3.26883$\times10^{-8}$$i$ & 1.13691$\times10^{-7}$ - 1.13691$\times10^{-7}i$ & 2.32256$\times10^{-7}$ - 2.32256$\times10^{-7}i$ \\
 
 0.8 & 3.88432$\times10^{-13}$ - 3.88434$\times10^{-13}$$i$ & 6.72785$\times10^{-13}$ - 6.72786$\times10^{-13}i$ & 8.68562$\times10^{-13}$ - 8.68563$\times10^{-13}i$ \\
 
 0.9 & 4.48074$\times10^{-13}$ - 4.48077$\times10^{-13}$$i$ & 7.76088$\times10^{-13}$ - 7.7609$\times10^{-13}i$ & 1.00193$\times10^{-12}$ - 1.00193$\times10^{-12}i$ \\
  
 1.0 & 5.05655$\times10^{-8}$ - 5.05658$\times10^{-8}$$i$ & 1.75869$\times10^{-7}$ - 1.7587$\times10^{-7}$$i$ & 3.59279$\times10^{-7}$ - 3.59279$\times10^{-7}i$ \\
   [0.2ex] 
 \hline \hline
\end{tabular}
\caption{\label{table5}Using the third--order WKB approximation, it is shown the \textit{quasinormal} frequencies for various values of mass $M$. Here, the multipole number is set to $l=0$.}
\end{center}
\end{table}

\begin{table}[!h]
\begin{center}
\begin{tabular}{c c c c} 
 \hline\hline
 $M$ & $\omega_{0}$ & $\omega_{1}$ & $\omega_{2}$  \\ [0.2ex] 
 \hline 
 0.1 & \text{Unstable} & \text{Unstable} & \text{Unstable}  \\ 
 
 0.2 & 0.043879 - 0.00511937$i$ & 0.043209 - 0.0154593$i$ & 0.0419976 - 0.0260389$i$ \\
 
 0.3 & 3.02554$\times10^{-14}$ - 1.65455$\times10^{-13}i$ & 8.23658$\times10^{-14}$ - 1.8233$\times10^{-13}i$  & 1.22805$\times10^{-13}$ - 2.03816$\times10^{-13}i$\\
 
 0.4 & 6.9521$\times10^{-14}$ - 2.40655$\times10^{-13}i$ & 1.73886$\times10^{-13}$ - 2.88649$\times10^{-13}i$  & 2.4743$\times10^{-13}$ - 3.38088$\times10^{-13}i$\\
 
 0.5 & 1.02399$\times10^{-13}$ - 2.95323$\times10^{-13}i$ & 2.45227$\times10^{-13}$ - 3.69955$\times10^{-13}i$ & 3.42973$\times10^{-13}$ - 4.40864$\times10^{-13}i$ \\
 
 0.6 & 1.31840$\times10^{-13}$ - 3.41425$\times10^{-13}i$ & 3.07025$\times10^{-13}$ - 4.39834$\times10^{-13}i$ & 4.25290$\times10^{-13}$ - 5.29208$\times10^{-13}i$ \\
 
 0.7 & 1.58961$\times10^{-13}$ - 3.82423$\times10^{-13}i$ & 3.62839$\times10^{-13}$ - 5.02624$\times10^{-13}i$ & 4.99423$\times10^{-13}$ - 6.08607$\times10^{-13}i$ \\
 
 0.8 & 1.84377$\times10^{-13}$ - 4.19946$\times10^{-13}i$ & 4.14448$\times10^{-13}$ - 5.6047$\times10^{-13}i$ & 5.67848$\times10^{-13}$ - 6.81771$\times10^{-13} i$ \\
 
 0.9 & 2.08474$10^{-13}$ - 4.54916$\times10^{-13}i$ & 4.62906$\times10^{-13}$ - 6.14629$\times10^{-13}i$ & 6.32015$\times10^{-13}$ - 7.50286$\times10^{-13}i$ \\
  
 1.0 & 2.31516$\times10^{-13}$ - 4.87918$\times10^{-13}i$ & 5.08899$\times10^{-13}$ - 6.65915$\times10^{-13}i$ & 6.92859$\times10^{-13}$ - 8.1518$\times10^{-13}i$ \\
   
  \\ [0.2ex] 
 \hline \hline
\end{tabular}
\caption{\label{table6}Using the third--order WKB approximation, it is shown the \textit{quasinormal} frequencies for various values of mass $M$. Here, the multipole number is set to $l=1$.}
\end{center}
\end{table}

\begin{table}[!h]
\begin{center}
\begin{tabular}{c c c c} 
 \hline\hline
 $M$ & $\omega_{0}$ & $\omega_{1}$ & $\omega_{2}$  \\ [0.2ex] 
 \hline 
 0.1 & \text{Unstable} & \text{Unstable} & \text{Unstable}   \\ 
 
 0.2 & 0.0755928 - 0.00511304$i$ & 0.0752001 - 0.0153732$i$ & 0.0744434 - 0.0257274$i$ \\
 
 0.3 & 3.02940$\times10^{-14}$ - 2.80618$\times10^{-13}i$ & 8.72488$\times10^{-14}$ - 2.92303$\times10^{-13}i$  & 1.36799$\times10^{-13}$ - 3.10713$\times10^{-13}i$\\
 
 0.4 & 7.04562$\times10^{-14}$ - 3.99604$\times10^{-13}i$ & 1.92813$\times10^{-13}$ - 4.38059$\times10^{-13}i$  & 2.88563$\times10^{-13}$ - 4.87840$\times10^{-13}i$\\
 
 0.5 & 1.04556$\times10^{-13}$ - 4.82169$\times10^{-13}i$ & 2.76936$\times10^{-13}$ - 5.46121$\times10^{-13}i$ & 4.05658$\times10^{-13}$ - 6.21380$\times10^{-13}i$ \\
 
 0.6 &1.35266$\times10^{-13}$ - 5.49558$\times10^{-13}i$ & 3.49924$\times10^{-13}$ - 6.37310$\times10^{-13}i$ & 5.05941$\times10^{-13}$ - 7.34638$\times10^{-13}i$ \\
 
 0.7 & 1.63600$\times10^{-13}$ - 6.07946$\times10^{-13}i$ & 4.15568$\times10^{-13}$ - 7.18004$\times10^{-13}i$ & 5.95481$\times10^{-13}$ - 8.35123$\times10^{-13}i$ \\
 
 0.8 & 1.90127$\times10^{-13}$ - 6.60219$\times10^{-13}i$ & 4.75878$\times10^{-13}$ - 7.91328$\times10^{-13}i$ & 6.77357$\times10^{-13}$ - 9.26582$\times10^{-13} i$ \\
 
 0.9 & 2.15216$\times10^{-13}$ - 7.08011$\times10^{-13}i$ & 5.32089$\times10^{-13}$ - 8.59116$\times10^{-13}i$ & 7.53413$\times10^{-13}$ - 1.01123$\times10^{-12}i$ \\
  
 1.0 & 2.39126$\times10^{-13}$ - 7.52352$\times10^{-13}i$ & 5.85027$\times10^{-13}$ - 9.22556$\times10^{-13}i$ & 8.24861$\times10^{-13}$ - 1.09053$\times10^{-12}i$ \\
   
  \\ [0.2ex] 
 \hline \hline
\end{tabular}
\caption{\label{table7}Using the third--order WKB approximation, it is shown the \textit{quasinormal} frequencies for various values of mass $M$. Here, the multipole number is set to $l=2$.}
\end{center}
\end{table}


\section{Conclusion}

In this study, we have focused on exploring the properties of a regular black hole within the framework of Verlinde's emergent gravity, with particular emphasis on the modified Simpson--Visser solution. Our analysis has revealed the existence of a single physical event horizon under certain conditions.

By examining the \textit{Hawking} temperature and heat capacity, we have identified several phase transitions in the black hole system. Furthermore, we have conducted an investigation of geodesic trajectories for photon--like particles, including the identification of critical orbits, i.e., known as photon spheres, and shadows. More so, we also have provided calculations of time delay and the deflection angle. Additionally, to validate our findings, we have incorporated an extra application within the context of high--energy astrophysical phenomena, specifically focusing on the deposition of neutrino energy.

Furthermore, we have employed third--order WKB approximations to study the behavior of quasinormal modes, offering valuable information about the damping and decay properties of scalar waves in the context of the black hole surrounded by dark matter. Within the context of the dark matter scenario, another aspect worthy of investigation is the influence of charge $Q$ on various aspects, such as the quasinormal modes, thermodynamic properties, shadows, and the photon sphere.

In the realm of future research, a fascinating direction for further investigation pertains to the analysis of quantum tunneling radiation and its associated backreaction effects, similar to the study encountered in Ref. \cite{silva2013quantum}. These and other ideas are now under development.



\section*{Acknowledgments}
\hspace{0.5cm}

The author would like to thank Fundação de Apoio à Pesquisa do Estado da Paraíba (FAPESQ) and Conselho Nacional de Desenvolvimento Cientíıfico e Tecnológico (CNPq)  -- [200486/2022-5] and [150891/2023-7] for the financial support. Most of the calculations were performed by using the \textit{Mathematica} software. Furthermore, the author extends gratitude to N. Heidari for invaluable suggestions and for providing the code used to compute the revised version of this manuscript. 

\section{Data Availability Statement}

Data Availability Statement: No Data associated in the manuscript


\bibliographystyle{ieeetr}
\bibliography{main}

\end{document}